\documentclass[12pt]{article}
\usepackage{amsmath,amssymb}
\usepackage{graphicx}
\newcommand{\eqdef}{\stackrel{\text{def}}{=}}
\newcommand{\n}{\nonumber\\}
\newcommand{\bm}{\boldsymbol}
\newcommand{\ignore}[1]{}
\numberwithin{equation}{section}
\newcommand{\Romannumeral}[1]{\uppercase\expandafter{\romannumeral#1}}

\newcommand{\II}{\text{\Romannumeral{2}}}

\allowdisplaybreaks[4]

\begin{document}

\baselineskip=20pt

\newfont{\elevenmib}{cmmib10 scaled\magstep1}
\newcommand{\preprint}{
    \begin{flushright}\normalsize \sf
     DPSU-17-2\\
   \end{flushright}}
\newcommand{\Title}[1]{{\baselineskip=26pt
   \begin{center} \Large \bf #1 \\ \ \\ \end{center}}}
\newcommand{\Author}{\begin{center}
   \large \bf Satoru Odake \end{center}}
\newcommand{\Address}{\begin{center}
     Faculty of Science, Shinshu University,\\
     Matsumoto 390-8621, Japan
   \end{center}}
\newcommand{\Accepted}[1]{\begin{center}
   {\large \sf #1}\\ \vspace{1mm}{\small \sf Accepted for Publication}
   \end{center}}

\preprint
\thispagestyle{empty}

\Title{Casoratian Identities\\ for the Discrete Orthogonal Polynomials\\
in Discrete Quantum Mechanics with Real Shifts}

\Author

\Address
\vspace{1cm}

\begin{abstract}
In our previous papers, the Wronskian identities for the Hermite, Laguerre
and Jacobi polynomials and the Casoratian identities for the Askey-Wilson
polynomial and its reduced form polynomials were presented.
These identities are naturally derived through quantum mechanical formulation
of the classical orthogonal polynomials; ordinary quantum mechanics for the
former and discrete quantum mechanics with pure imaginary shifts for the latter.
In this paper we present the corresponding identities for the discrete
quantum mechanics with real shifts.
Infinitely many Casoratian identities for the $q$-Racah polynomial and its
reduced form polynomials are obtained.
\end{abstract}

\section{Introduction}
\label{intro}

New types of orthogonal polynomials, the exceptional and multi-indexed
orthogonal polynomials $\{P_{\mathcal{D},n}(\eta)|n\in\mathbb{Z}_{\geq 0}\}$,
have made remarkable progress in the theory of orthogonal polynomials and
exactly solvable quantum mechanical models \cite{gkm08}--\cite{os34}.
Our approach to orthogonal polynomials is based on the quantum mechanical
formulations: ordinary quantum mechanics (oQM), discrete quantum mechanics
with pure imaginary shifts (idQM) \cite{os13}--\cite{os24} and discrete
quantum mechanics with real shifts (rdQM) \cite{os12}--\cite{os34}.
For oQM, Schr\"odinger equations are second order differential equations.
In discrete quantum mechanics, they are replaced by second order difference
equations with a continuous variable for idQM or a discrete variable for rdQM.
The Askey scheme of the (basic) hypergeometric orthogonal polynomials
\cite{kls} is well matched to these quantum mechanical formulations:
the Jacobi polynomial etc.\ in oQM, the Askey-Wilson polynomial etc.\ in
idQM and the $q$-Racah polynomial etc.\ in rdQM.
{}From the exactly solvable quantum mechanical systems described by the
classical orthogonal polynomials in the Askey scheme, we can obtain new
exactly solvable quantum mechanical systems and various exceptional
orthogonal polynomials with multi-indices by the multi-step Darboux
transformations with appropriate seed solutions.
One characteristic feature of these new types polynomials is the missing
degrees.
We distinguish the following two cases;
the set of missing degrees $\mathcal{I}=\mathbb{Z}_{\geq 0}\backslash
\{\text{deg}\,\mathcal{P}_n|n\in\mathbb{Z}_{\geq 0}\}$ is
case (1): $\mathcal{I}=\{0,1,\ldots,\ell-1\}$, or
case (2): $\mathcal{I}\neq\{0,1,\ldots,\ell-1\}$, where $\ell$ is a positive
integer. The situation of case (1) is called stable in \cite{gkm11}.
When the virtual state wavefunctions are used as seed solutions, the
deformed systems are exactly iso-spectral to the original system and
the case (1) multi-indexed orthogonal polynomials are obtained
\cite{os25,os27,os26}.
When the eigenstate and/or pseudo virtual state wavefunctions are used as seed
solutions, the deformed systems are almost iso-spectral to the original
system but some states corresponding to the seed solutions are deleted or
added, respectively, and the case (2) multi-indexed orthogonal polynomials
are obtained \cite{os29,os30}.

For oQM and idQM, the deformed systems obtained by $M$-step Darboux
transformations in terms of pseudo virtual state wavefunctions (with degrees
specified by $\mathcal{D}$) are equivalent to those obtained by multi-step
Darboux transformations in terms of eigenstate wavefunctions (with degrees
specified by $\bar{\mathcal D}$) with shifted parameters \cite{os29,os30}.
The deformed system is characterized by the denominator polynomial
$\Xi_{\mathcal{D}}(\eta;\bm{\lambda})$ and the above equivalence is
based on the proportionality of the denominator polynomials for each
deformed system,
\begin{equation}
  \Xi_{\mathcal{D}}(\eta;\bm{\lambda})\propto
  \bar{\Xi}_{\bar{\mathcal{D}}}(\eta;\bar{\bm{\lambda}}),
  \label{Xi=bXi}
\end{equation}
see \cite{os29,os30} for details.
Here we explain necessary notation only.
Two sets $\mathcal{D}$ and $\bar{\mathcal{D}}$ are defined for
positive integers $M$ and $\mathcal{N}$
($\bm{\lambda}$ is a set of parameters and $\bm{\delta}$ is its shift.):
\begin{align}
  &\mathcal{D}=\{d_1,d_2,\ldots,d_M\}
  \ \ (d_j\in\mathbb{Z}_{\geq 0}\, :\text{mutually distinct}),\n
  &\mathcal{N}\geq\max(\mathcal{D}),\quad \bar{d}_j\eqdef\mathcal{N}-d_j,\quad
  \bar{\bm{\lambda}}\eqdef\bm{\lambda}-(\mathcal{N}+1)\bm{\delta},\quad
  \bar{\mathcal{N}}\eqdef\mathcal{N}+1-M,
  \label{Dbar}\\
  &\bar{\mathcal{D}}\eqdef
  \{0,1,\ldots,\mathcal{N}\}
  \backslash\{\bar{d}_1,\bar{d}_2,\ldots,\bar{d}_M\}
  \eqdef\{e_1,e_2,\ldots,e_{\bar{\mathcal{N}}}\}.\nonumber
\end{align}
(Exactly speaking, $\mathcal{D}$ and $\bar{\mathcal{D}}$ should be treated
as ordered sets.
By changing the order of $d_j$'s, $\Xi_{\mathcal{D}}(\eta;\bm{\lambda})$
changes its overall sign.
For the proportional relation \eqref{Xi=bXi}, however, such an overall sign
change does not matter.)
The proportional relation \eqref{Xi=bXi} gives the Wronskian identity
for oQM and the Casoratian identity for idQM.
We write down the Casoratian identities for the Askey-Wilson polynomial:
\begin{equation}
  \varphi_M(x)^{-1}\,
  \text{W}_{\gamma}[\check{\xi}_{d_1},\check{\xi}_{d_2},\ldots,
  \check{\xi}_{d_M}](x;\bm{\lambda})
  \propto
  \varphi_{\bar{\mathcal{N}}}(x)^{-1}\,
  \text{W}_{\gamma}[\check{P}_{e_1},\check{P}_{e_2},\ldots,
  \check{P}_{e_{\bar{\mathcal{N}}}}](x;\bar{\bm{\lambda}}).
  \label{casoidAW}
\end{equation}
(See Appendix \ref{app:AW} for definitions of
$\check{\xi}_{\text{v}}(x;\bm{\lambda})$, $\check{P}_n(x;\bm{\lambda})$
and $\varphi_M(x)$.)
Here the Casorati determinant (Casoratian) of a set of $n$ functions
$\{f_j(x)\}$ for idQM is defined by
\begin{equation}
  \text{W}_{\gamma}[f_1,f_2,\ldots,f_n](x)
  \eqdef i^{\frac12n(n-1)}
  \det\Bigl(f_k\bigl(x+i(\tfrac{n+1}{2}-j)\gamma\bigr)\Bigr)_{1\leq j,k\leq n},
  \label{idQM:Wdef}
\end{equation}
(for $n=0$, we set $\text{W}_{\gamma}[\cdot](x)=1$) and $\gamma=\log q$ for
the Askey-Wilson case.
Based on the discrete symmetry of the system, the pseudo virtual state
polynomial $\check{\xi}_{\text{v}}(x;\bm{\lambda})$ is obtained from the
eigenpolynomial $\check{P}_n(x;\bm{\lambda})$ by twisting parameters.
The Casoratian identities \eqref{casoidAW} represent the relation between
Casoratians of the orthogonal polynomials with twisted and shifted parameters,
and display the duality between the state adding and deleting Darboux
transformations.
The shape invariant properties of the original systems play an important role.

In this paper we consider the Casoratian identities for discrete orthogonal
polynomials appearing in rdQM.
A natural way to obtain them is the following; (a) Define the pseudo virtual
state vectors by using discrete symmetries of the system, (b) Deform the
system by multi-step Darboux transformations in terms of the pseudo virtual
state vectors, (c) Compare it with the deformed system obtained by multi-step
Darboux transformations in terms of the eigenvectors with shifted parameters.
We present one-step Darboux transformation in terms of the pseudo virtual
state vector by taking $q$-Racah case as an example in Appendix\,\ref{app:pvsv}.
However, calculation for multi-step cases is rather complicated.
So, instead of the natural method mentioned above, we use a `shortcut'
method in this paper.
We derive the Casoratian identities for the $q$-Racah polynomial
\eqref{casoidqR} from those for the Askey-Wilson polynomial \eqref{casoidAW}
by using the relation between the $q$-Racah and Askey-Wilson polynomials.
The discrete orthogonal polynomials in the Askey scheme are obtained from
the $q$-Racah polynomial in appropriate limits. The Casoratian identities
for those reduced form polynomials can be obtained from those for the
$q$-Racah polynomial in the same limits.

The Casoratian identities imply equivalences between the deformed systems
obtained by multi-step Darboux transformations in terms of pseudo virtual
state vectors and those in terms of eigenvectors with shifted parameters.
For each (exactly solvable) rdQM system, we can construct the (exactly
solvable) birth and death process \cite{s09}, which is a stationary Markov
chain. The Casoratian identities provide equivalence among such birth
and death processes.

Similar Casoratian identities were studied by Curbera and Dur\'an \cite{cd16}.
Their method is different from ours and the identities are presented
for the Charlier, Meixner, Krawtchouk and Hahn polynomials only, which have the
sinusoidal coordinate $\eta(x)=x$.
In our method various polynomials having five types of sinusoidal coordinates
\cite{os12} $\eta(x)=x,x(x+d),1-q^x,q^{-x}-1,(q^{-x}-1)(1-dq^x)$ are covered.

This paper is organized as follows.
In section \ref{sec:rdQM} we recapitulate the discrete quantum mechanics
with real shifts.
Section \ref{sec:Casoid} is the main part of this paper.
After presenting the data for the original ($q$-)Racah systems in
\S\,\ref{sec:original}, we discuss their discrete symmetries and present
the pseudo virtual state polynomials by using the twist operations in
\S\,\ref{sec:discsym}.
The Casoratian identities for the ($q$-)Racah polynomials are derived
starting from those for the Askey-Wilson polynomial in \S\,\ref{sec:casoidqR}.
In section \ref{sec:Casoidred} the Casoratian identities for the reduced
form polynomials are presented.
Section \ref{sec:summary} is for a summary and comments.
In Appendix \ref{app:data} some necessary data of orthogonal polynomials are
presented.
In Appendix\,\ref{app:pvsv} the pseudo virtual state vectors and one-step
Darboux transformation are discussed by taking the $q$-Racah system as an
example.

\section{Discrete Quantum Mechanics with Real Shifts}
\label{sec:rdQM}

In this section we recapitulate the discrete quantum mechanics with real
shifts (rdQM) developed in \cite{os12,os34}. 

The Hamiltonian of rdQM $\mathcal{H}=(\mathcal{H}_{x,y})$ is a tri-diagonal
real symmetric (Jacobi) matrix and its rows and columns are indexed by
integers $x$ and $y$, which take values in $\{0,1,\ldots,N\}$ (finite) or
$\mathbb{Z}_{\geq 0}$ (semi-infinite),
\begin{equation}
  \mathcal{H}_{x,y}\eqdef
  -\sqrt{B(x)D(x+1)}\,\delta_{x+1,y}-\sqrt{B(x-1)D(x)}\,\delta_{x-1,y}
  +\bigl(B(x)+D(x)\bigr)\delta_{x,y}.
  \label{rdQM:Hxy}
\end{equation}
The potential functions $B(x)$ and $D(x)$ are real and positive but vanish
at the boundary, $D(0)=0$ for both cases and $B(N)=0$ for a finite case.
In this paper we consider the case that these $B(x)$ and $D(x)$ are rational
functions of $x$ or $q^x$ ($0<q<1$).
For simplicity in notation, we write the matrix $\mathcal{H}$ as follows:
\begin{align}
  &\mathcal{H}=-\sqrt{B(x)D(x+1)}\,e^{\partial}
  -\sqrt{B(x-1)D(x)}\,e^{-\partial}+B(x)+D(x)\n
  &\phantom{\mathcal{H}}=-\sqrt{B(x)}\,e^{\partial}\sqrt{D(x)}
  -\sqrt{D(x)}\,e^{-\partial}\sqrt{B(x)}+B(x)+D(x),
  \label{rdQM:H}
\end{align}
where matrices $e^{\pm\partial}$ are
\begin{equation}
  e^{\pm\partial}=((e^{\pm\partial})_{x,y}),\quad
  (e^{\pm\partial})_{x,y}\eqdef\delta_{x\pm 1,y},\quad
  (e^{\partial})^{\dagger}=e^{-\partial},
  \label{partdef}
\end{equation}
and we suppress the unit matrix $\bm{1}=(\delta_{x,y})$ :
$\bigl(B(x)+D(x)\bigr)\bm{1}$ in \eqref{rdQM:H}.
The notation $f(x)Ag(x)$, where $f(x)$ and $g(x)$ are functions of $x$ and
$A$ is a matrix $A=(A_{x,y})$, stands for a matrix whose $(x,y)$-element
is $f(x)A_{x,y}g(y)$.
Note that the matrices $e^{\partial}$ and $e^{-\partial}$ are not inverse
to each other: $e^{\pm\partial}e^{\mp\partial}\neq\bm{1}$ for a finite system
and $e^{-\partial}e^{\partial}\neq\bm{1}$ for a semi-infinite system.
This Hamiltonian can be expressed in a factorized form:
\begin{equation}
  \mathcal{H}=\mathcal{A}^{\dagger}\mathcal{A},\quad
  \mathcal{A}\eqdef\sqrt{B(x)}-e^{\partial}\sqrt{D(x)},\quad
  \mathcal{A}^{\dagger}=\sqrt{B(x)}-\sqrt{D(x)}\,e^{-\partial}.
  \label{factor}
\end{equation}

The Schr\"odinger equation is the eigenvalue problem for the hermitian
matrix $\mathcal{H}$,
\begin{equation}
  \mathcal{H}\phi_n(x)=\mathcal{E}_n\phi_n(x)\quad
  (n=0,1,2,\ldots),\quad
  0=\mathcal{E}_0<\mathcal{E}_1<\mathcal{E}_2<\cdots,
  \label{Scheq}
\end{equation}
($n=0,1,\ldots,N$ for a finite case).
The ground state eigenvector $\phi_0(x)$, which is characterized by
$\mathcal{A}\phi_0(x)=0$, is chosen as
\begin{equation}
  \phi_0(x)=\prod_{y=0}^{x-1}\sqrt{\frac{B(y)}{D(y+1)}}\,>0.
  \label{phi0}
\end{equation}
We use the convention: $\prod\limits_{k=n}^{n-1}*=1$, which means the
normalization $\phi_0(0)=1$.
Remark that the boundary condition $D(0)=0$ and $B(N)=0$ for a finite case
is important for the zero mode equation $\mathcal{H}\phi_0(x)=0$,
cf.\,\eqref{H'phi'0}.
For the original systems (not deformed one) considered in this paper,
the eigenvectors have the following factorized form
\begin{equation}
  \phi_n(x)=\phi_0(x)\check{P}_n(x),\quad
  \check{P}_n(x)\eqdef P_n\bigl(\eta(x)\bigr).
\end{equation}
Here $P_n(\eta)$ is a polynomial of degree $n$ in $\eta$, and
the sinusoidal coordinate $\eta(x)$ is one of the following \cite{os12}:
$\eta(x)=x,\epsilon'x(x+d),1-q^x,q^{-x}-1,\epsilon'(q^{-x}-1)(1-dq^x)$,
($\epsilon'=\pm 1$), which satisfy the boundary condition $\eta(0)=0$.
We adopt the universal normalization condition \cite{os12,os34} as
\begin{equation}
  P_n(0)=1\ \bigl(\Leftrightarrow \check{P}_n(0)=1\bigr).
  \label{Pzero}
\end{equation}
This $\check{P}_n(x)$ is the eigenvector of the similarity transformed
Hamiltonian $\widetilde{\mathcal{H}}$,
\begin{align}
  &\widetilde{\mathcal{H}}
  \eqdef\phi_0(x)^{-1}\circ\mathcal{H}\circ\phi_0(x)
  =B(x)(1-e^{\partial})+D(x)(1-e^{-\partial}),
  \label{Ht}\\
  &\widetilde{\mathcal{H}}\check{P}_n(x)=\mathcal{E}_n\check{P}_n(x).
  \label{HtcP=}
\end{align}
Explicitly, \eqref{HtcP=} is the difference equation for $\check{P}_n$,
\begin{equation}
  B(x)\bigl(\check{P}_n(x)-\check{P}_n(x+1)\bigr)
  +D(x)\bigl(\check{P}_n(x)-\check{P}_n(x-1)\bigr)
  =\mathcal{E}_n\check{P}_n(x).
  \label{diffeqcPn}
\end{equation}
Since $P_n$ is a polynomial, $\check{P}_n(x)$ is defined for any
$x\in\mathbb{R}$ and the difference equation \eqref{diffeqcPn}
is also valid for $x\in\mathbb{R}$.
The eigenvectors are mutually orthogonal ($d_n>0$):
\begin{equation}
  (\phi_n,\phi_m)\eqdef\sum_{x=0}^{x_{\text{max}}}\phi_n(x)\phi_m(x)
  =\sum_{x=0}^{x_{\text{max}}}\phi_0(x)^2\check{P}_n(x)\check{P}_m(x)
  =\frac{1}{d_n^2}\delta_{nm},
\end{equation}
where $x_{\text{max}}=N$ for a finite case, $\infty$ for a semi-infinite case.
(Although this notation $d_n$ conflicts with the notation of the label of the
pseudo virtual vector $d_j$ in \eqref{Dbar}, we think this does not cause any
confusion because the former appears as $\frac{1}{d_n^2}\,\delta_{nm}$.)

If we find functions $B'(x)$ and $D'(x)$ satisfying
\begin{align}
  B(x)D(x+1)&=\alpha^2B'(x)D'(x+1),\quad\alpha\neq 0,
  \label{BD=B'D'}\\
  B(x)+D(x)&=\alpha\bigl(B'(x)+D'(x)\bigr)+\alpha',
  \label{B+D=B'+D'}
\end{align}
where $\alpha$ and $\alpha'$ are real constants, we obtain the following
relation:
\begin{equation}
  \mathcal{H}=\alpha\mathcal{H}'+\alpha',
  \label{H=aH'+a'}
\end{equation}
where $\mathcal{H}'=(\mathcal{H}'_{x,y})_{0\leq x,y\leq x_{\text{max}}}$ is
given by
\begin{equation}
  \mathcal{H}'=-\frac{|\alpha|}{\alpha}\sqrt{B'(x)D'(x+1)}\,e^{\partial}
  -\frac{|\alpha|}{\alpha}\sqrt{B'(x-1)D'(x)}\,e^{-\partial}+B'(x)+D'(x).
  \label{H'}
\end{equation}

For concrete examples, various quantities depend on a set of parameters
$\bm{\lambda}=(\lambda_1,\lambda_2,\ldots)$ and $q$.
The parameter $q$ is $0<q<1$ and $q^{\bm{\lambda}}$ stands for
$q^{(\lambda_1,\lambda_2,\ldots)}=(q^{\lambda_1},q^{\lambda_2},\ldots)$.
The $\bm{\lambda}$-dependence is expressed like,
$\mathcal{H}=\mathcal{H}(\bm{\lambda})$,
$\mathcal{A}=\mathcal{A}(\bm{\lambda})$,
$\mathcal{E}_n=\mathcal{E}_n(\bm{\lambda})$,
$B(x)=B(x;\bm{\lambda})$,
$\phi_n(x)=\phi_n(x;\bm{\lambda})$,
$\check{P}_n(x)=\check{P}_n(x;\bm{\lambda})
=P_n\bigl(\eta(x;\bm{\lambda});\bm{\lambda}\bigr)$, etc.
If needed, the $q$-dependence is also expressed like,
$\mathcal{E}_n=\mathcal{E}_n(\bm{\lambda};q)$,
$B(x)=B(x;\bm{\lambda};q)$,
$\check{P}_n(x)=\check{P}_n(x;\bm{\lambda};q)
=P_n\bigl(\eta(x;\bm{\lambda};q);\bm{\lambda};q\bigr)$, etc.

The original systems in this paper are shape invariant
\cite{os12} and they satisfy the relation,
\begin{equation}
  \mathcal{A}(\bm{\lambda})\mathcal{A}(\bm{\lambda})^{\dagger}
  =\kappa\mathcal{A}(\bm{\lambda}+\bm{\delta})^{\dagger}
  \mathcal{A}(\bm{\lambda}+\bm{\delta})+\mathcal{E}_1(\bm{\lambda}),
  \quad\kappa>0,
\end{equation}
which is a sufficient condition for exact solvability.
The auxiliary functions $\varphi(x;\bm{\lambda})$ \cite{os12} and
$\varphi_M(x;\bm{\lambda})$ \cite{os22} are defined by
\begin{align}
  \varphi(x;\bm{\lambda})&\eqdef
  \frac{\eta(x+1;\bm{\lambda})-\eta(x;\bm{\lambda})}{\eta(1;\bm{\lambda})},
  \label{varphi}\\
  \varphi_M(x;\bm{\lambda})&\eqdef\prod_{1\leq j<k\leq M}
  \frac{\eta(x+k-1;\bm{\lambda})-\eta(x+j-1;\bm{\lambda})}
  {\eta(k-j;\bm{\lambda})}\n
  &=\prod_{1\leq j<k\leq M}
  \varphi\bigl(x+j-1;\bm{\lambda}+(k-j-1)\bm{\delta}\bigr),
  \label{varphiM}
\end{align}
and $\varphi_0(x;\bm{\lambda})=\varphi_1(x;\bm{\lambda})=1$.

For the orthogonal polynomials with Jackson integral measures such as the
big $q$-Jacobi polynomial, the two component formulation is needed,
see \cite{os34}.

The symbols $(a)_n$ and $(a;q)_n$ are ($q$-)shifted factorials
(($q$-)Pochhammer symbols) \cite{kls}. They are defined for a non-negative
integer $n$ by $(a)_n=\prod_{j=0}^{n-1}(a+j)$ and
$(a;q)_n=\prod_{j=0}^{n-1}(1-aq^j)$,
which are extended to a real $n$ by
\begin{equation}
  (a)_n=\frac{\Gamma(a+n)}{\Gamma(a)},\quad
  (a;q)_n=\frac{(a;q)_{\infty}}{(aq^n;q)_{\infty}}.
  \label{qPoch}
\end{equation}
Note that, for $a\neq 0$ and $n\in\mathbb{Z}_{\geq 0}$,
\begin{equation}
  (a;q^{-1})_n=(-a)^nq^{-\frac12n(n-1)}(a^{-1};q)_n.
  \label{qPochid}
\end{equation}
The hypergeometric series ${}_rF_s$ and
the basic hypergeometric series ${}_r\phi_s$ are
\begin{align}
  {}_rF_s\Bigl(\genfrac{}{}{0pt}{}{a_1,\,\cdots,a_r}{b_1,\,\cdots,b_s}
  \Bigm|z\Bigr)
  &\eqdef\sum_{n=0}^{\infty}
  \frac{(a_1,\,\cdots,a_r)_n}{(b_1,\,\cdots,b_s)_n}\frac{z^n}{n!},\\
  {}_r\phi_s\Bigl(
  \genfrac{}{}{0pt}{}{a_1,\,\cdots,a_r}{b_1,\,\cdots,b_s}
  \Bigm|q\,;z\Bigr)
  &\eqdef\sum_{n=0}^{\infty}
  \frac{(a_1,\,\cdots,a_r;q)_n}{(b_1,\,\cdots,b_s;q)_n}
  (-1)^{(1+s-r)n}q^{(1+s-r)\frac12n(n-1)}\frac{z^n}{(q;q)_n},
\end{align}
where $(a_1,\,\cdots,a_r)_n\eqdef\prod_{k=1}^r(a_k)_n$
and $(a_1,\,\cdots,a_r;q)_n\eqdef\prod_{k=1}^r(a_k;q)_n$.

The Casorati determinant (Casoratian) of a set of $n$ functions $\{f_j(x)\}$
for rdQM is defined by
\begin{equation}
  \text{W}_{\text{C}}[f_1,f_2,\ldots,f_n](x)
  \eqdef\det\Bigl(f_k(x+j-1)\Bigr)_{1\leq j,k\leq n},
  \label{rdQM:Wdef}
\end{equation}
(for $n=0$, we set $\text{W}_{\text{C}}[\cdot](x)=1$).

For well-defined quantum systems, the range of parameters $\bm{\lambda}$
must be chosen such that the Hamiltonian is real symmetric.
On the other hand, our main purpose of this paper is to obtain the Casoratian
identities \eqref{casoidqR}. Both sides of \eqref{casoidqR} are polynomials
in $x$ or Laurent polynomials in $q^x$ and \eqref{casoidqR} hold for any
parameter range (except for the zeros of the denominators). 
So we do not bother about the range of parameters, except for
Appendix\,\ref{app:pvsv}.

\section{Casoratian Identities of the ($q$-)Racah Polynomials}
\label{sec:Casoid}

In this section we consider rdQM whose eigenvectors are described by the
($q$-)Racah polynomials.
After discussing some discrete symmetries and pseudo virtual state polynomials,
the Casoratian identities of the ($q$-)Racah polynomials are presented.

\subsection{Original systems}
\label{sec:original}

Let us consider the Racah(R) and $q$-Racah($q$R) systems \cite{os12}.
Although there are four possible parameter choices indexed by
$(\epsilon,\epsilon')=(\pm 1,\pm 1)$ in general, we restrict ourselves
to the $(\epsilon,\epsilon')=(1,1)$ case for simplicity of presentation.
The set of parameters $\bm{\lambda}=(\lambda_1,\lambda_2,\lambda_3,\lambda_4)$,
its shift $\bm{\delta}$ and $\kappa$ are
\begin{align}
  \text{R}:\ \ &\bm{\lambda\,}=(a,b,c,d),\quad \bm{\delta}=(1,1,1,1),
  \quad\kappa=1,\\
  \text{$q$R}:\ \ &q^{\bm{\lambda}}=(a,b,c,d),\quad \bm{\delta}=(1,1,1,1),
  \quad\kappa=q^{-1},
\end{align}
and we take
\begin{equation}
  \lambda_3=-N,\quad\text{namely}\ \ c=\left\{
  \begin{array}{ll}
  -N&:\text{R}\\
  q^{-N}&:\text{$q$R}
  \end{array}\right..
\end{equation}
We list the fundamental data:
\begin{align}
  &B(x;\bm{\lambda})=
  \left\{
  \begin{array}{ll}
  {\displaystyle
  -\frac{(x+a)(x+b)(x+c)(x+d)}{(2x+d)(2x+1+d)}}&:\text{R}\\[8pt]
  {\displaystyle-\frac{(1-aq^x)(1-bq^x)(1-cq^x)(1-dq^x)}
  {(1-dq^{2x})(1-dq^{2x+1})}}&:\text{$q$R}
  \end{array}\right.,\\
  &D(x;\bm{\lambda})=
  \left\{
  \begin{array}{ll}
  {\displaystyle
  -\frac{(x+d-a)(x+d-b)(x+d-c)x}{(2x-1+d)(2x+d)}}&:\text{R}\\[8pt]
  {\displaystyle-\tilde{d}\,
  \frac{(1-a^{-1}dq^x)(1-b^{-1}dq^x)(1-c^{-1}dq^x)(1-q^x)}
  {(1-dq^{2x-1})(1-dq^{2x})}}&:\text{$q$R}
  \end{array}\right.,\\
  &\mathcal{E}_n(\bm{\lambda})=
  \left\{
  \begin{array}{ll}
  n(n+\tilde{d})&:\text{R}\\[2pt]
  (q^{-n}-1)(1-\tilde{d}q^n)&:\text{$q$R}
  \end{array}\right.\!,\quad
  \eta(x;\bm{\lambda})=
  \left\{
  \begin{array}{ll}
  x(x+d)&:\text{R}\\[2pt]
  (q^{-x}-1)(1-dq^x)&:\text{$q$R}
  \end{array}\right.,
  \label{EnqR}\\
  &\varphi(x;\bm{\lambda})=
  \left\{
  \begin{array}{ll}
  {\displaystyle\frac{2x+d+1}{d+1}}&:\text{R}\\[6pt]
  {\displaystyle\frac{q^{-x}-dq^{x+1}}{1-dq}}&:\text{$q$R}
  \end{array}\right.,\quad
  \tilde{d}\eqdef
  \left\{
  \begin{array}{ll}
  a+b+c-d-1&:\text{R}\\[2pt]
  abcd^{-1}q^{-1}&:\text{$q$R}
  \end{array}\right.,\\
  &\check{P}_n(x;\bm{\lambda})
  =P_n\bigl(\eta(x;\bm{\lambda});\bm{\lambda}\bigr)
  =\left\{
  \begin{array}{ll}
  {\displaystyle
  {}_4F_3\Bigl(
  \genfrac{}{}{0pt}{}{-n,\,n+\tilde{d},\,-x,\,x+d}
  {a,\,b,\,c}\Bigm|1\Bigr)}&:\text{R}\\[8pt]
  {\displaystyle
  {}_4\phi_3\Bigl(
  \genfrac{}{}{0pt}{}{q^{-n},\,\tilde{d}q^n,\,q^{-x},\,dq^x}
  {a,\,b,\,c}\Bigm|q\,;q\Bigr)}&:\text{$q$R}
  \end{array}\right.\n
  &\phantom{\check{P}_n(x;\bm{\lambda})
  =P_n\bigl(\eta(x;\bm{\lambda});\bm{\lambda}\bigr)}
  =\left\{
  \begin{array}{ll}
  {\displaystyle
  R_n\bigl(\eta(x;\bm{\lambda});a-1,\tilde{d}-a,c-1,d-c\bigr)}
  &:\text{R}\\[2pt]
  {\displaystyle
  R_n\bigl(1+d+\eta(x;\bm{\lambda});
  aq^{-1},\tilde{d}a^{-1},cq^{-1},dc^{-1}|q\bigr)}&:\text{$q$R}
  \end{array}\right.,\\[4pt]
  &\phi_0(x;\bm{\lambda})^2=\left\{
  \begin{array}{ll}
  {\displaystyle
  \frac{(a,b,c,d)_x}{(d-a+1,d-b+1,d-c+1,1)_x}\,\frac{2x+d}{d}}
  &:\text{R}\\[12pt]
  {\displaystyle
  \frac{(a,b,c,d\,;q)_x}
  {(a^{-1}dq,b^{-1}dq,c^{-1}dq,q\,;q)_x\,\tilde{d}^x}\,\frac{1-dq^{2x}}{1-d}}
  &:\text{$q$R}
  \end{array}\right.,
  \label{phi0qR}
\end{align}
where $R_n\bigl(x(x+\gamma+\delta+1);\alpha,\beta,\gamma,\delta\bigr)$ and
$R_n(q^{-x}+\gamma\delta q^{x+1};\alpha,\beta,\gamma,\delta|q)$ are the Racah
and $q$-Racah polynomials in the conventional parametrization \cite{kls},
respectively.
Our parametrization respects the correspondence between the ($q$-)Racah and
(Askey-)Wilson polynomials, and symmetries in $(a,b,c,d)$ are transparent.
The expressions of the right hand sides of $\mathcal{E}_n(\bm{\lambda})$,
$\eta(x;\bm{\lambda})$, $\varphi(x;\bm{\lambda})$,
$\check{P}_n(x;\bm{\lambda})$ and $\phi_0(x;\bm{\lambda})^2$ in
\eqref{EnqR}--\eqref{phi0qR} are also valid for real $n$ or real $x$, and
we regard $\mathcal{E}_n(\bm{\lambda})$ etc.\ as functions of real $n$ or
real $x$ by \eqref{EnqR}--\eqref{phi0qR}.
Note that $\phi_0(x;\bm{\lambda})^2=0$ for
$x\in\mathbb{Z}\backslash\{0,1,\ldots,N\}$ due to the factor
$(c)_x/(1)_x$ or $(c;q)_x/(q;q)_x$.
The R and $q$R systems are invariant under the exchange $a\leftrightarrow b$.

The potential functions $B(x;\bm{\lambda})$ and $D(x;\bm{\lambda})$ have
the following symmetries: 
\begin{align}
  \text{R,\,$q$R}:&
  \ \ B(N-x;\bm{\lambda}')=D(x;\bm{\lambda}),\quad
  D(N-x;\bm{\lambda}')=B(x;\bm{\lambda}),\n
  & \ \ \bm{\lambda}'=
  (\lambda_1+\lambda_3-\lambda_4,\lambda_2+\lambda_3-\lambda_4,
  \lambda_3,2\lambda_3-\lambda_4),
  \label{sym0}\\
  \text{$q$R}:&
  \ \ B(x;\bm{\lambda};q^{-1})=\tilde{d}^{-1}B(x;\bm{\lambda};q),\quad
  D(x;\bm{\lambda};q^{-1})=\tilde{d}^{-1}D(x;\bm{\lambda};q).
  \label{sym0t}
\end{align}
Corresponding to \eqref{sym0}, the identities (1.7.6) and (1.13.26) in
\cite{kls} give the relations
\begin{equation}
  \check{P}_n(N-x;\bm{\lambda}')=\check{P}_n(x;\bm{\lambda})\times\left\{
  \begin{array}{ll}
  {\displaystyle\frac{(a,b)_n}{(1-a+\tilde{d},1-b+\tilde{d})_n}}
  &:\text{R}\\[10pt]
  {\displaystyle\frac{c^n(a,b;q)_n}
  {d^n(a^{-1}\tilde{d}q,b^{-1}\tilde{d}q;q)_n}}
  &:\text{$q$R}
  \end{array}
  \right..
\end{equation}
Corresponding to \eqref{sym0t}, the $q$R polynomial is invariant under
$q\to q^{-1}$,
\begin{equation}
  \check{P}_n(x;\bm{\lambda};q^{-1})=\check{P}_n(x;\bm{\lambda};q),
  \label{qRq^{-1}}
\end{equation}
which is shown by \eqref{qPochid}.
In the conventional notation, this \eqref{qRq^{-1}} is written as
\begin{equation*}
  R_n(q^x+\gamma^{-1}\delta^{-1}q^{-x-1};
  \alpha^{-1},\beta^{-1},\gamma^{-1},\delta^{-1}|q^{-1})
  =R_n(q^{-x}+\gamma\delta q^{x+1};\alpha,\beta,\gamma,\delta|q).
\end{equation*}

\subsection{Discrete symmetries}
\label{sec:discsym}

Let us consider the twist operation $\mathfrak{t}$, which is an involution
acting on $x$, $\bm{\lambda}$ and $q$,
\begin{equation}
  \mathfrak{t}(x,\bm{\lambda},q)
  =\bigl(\mathfrak{t}(x),\mathfrak{t}(\bm{\lambda}),\mathfrak{t}(q)\bigr),\quad
  \mathfrak{t}^2=\text{id}.
  \label{twist}
\end{equation}
For R system, the $q$-part should be ignored.
We present twist operations (\romannumeral1)-(\romannumeral2) and
($\widetilde{\text{\romannumeral1}}$)--($\widetilde{\text{\romannumeral2}}$),
which lead to the pseudo virtual state vectors explained in
Appendix\,\ref{app:pvsv}.

First let us define twist (\romannumeral2) as follows:
\begin{align}
  (\text{\romannumeral2}):&
  \ \ \mathfrak{t}(x)\eqdef x+\lambda_3-1,
  \ \ \mathfrak{t}(\bm{\lambda})
  \eqdef(2-\lambda_1-\lambda_3+\lambda_4,2-\lambda_2-\lambda_3+\lambda_4,
  2-\lambda_3,2-2\lambda_3+\lambda_4),\n
  &\ \ \mathfrak{t}(q)\eqdef q.
  \label{twist(ii)}
\end{align}
By using this twist operation, the functions $B'(x)$ and $D'(x)$ satisfying
\eqref{BD=B'D'}--\eqref{B+D=B'+D'} are obtained:
\begin{align}
  (\text{\romannumeral2}):&
  \ \ B'(x;\bm{\lambda})\eqdef B\bigl(x-N-1;\mathfrak{t}(\bm{\lambda})\bigr),
  \ \ D'(x;\bm{\lambda})\eqdef D\bigl(x-N-1;\mathfrak{t}(\bm{\lambda})\bigr),
  \label{B'D'(ii)}\\
  &\ \ \alpha(\bm{\lambda})=\left\{
  \begin{array}{ll}
  1&:\text{R}\\
  \tilde{d}q^{-1}&:\text{$q$R}
  \end{array}
  \right.,\quad
  \alpha'(\bm{\lambda})=\left\{
  \begin{array}{ll}
  -(\tilde{d}-1)&:\text{R}\\
  -(1-q)(1-\tilde{d}q^{-1})&:\text{$q$R}
  \end{array}
  \right..
  \label{alal'(ii)}
\end{align}
Explicitly $B'(x)$ and $D'(x)$ are
\begin{align}
  (\text{\romannumeral2}):&
  \ \ B'(x;\bm{\lambda})=D(x+1;\bm{\lambda})\times\left\{
  \begin{array}{ll}
  {\displaystyle\frac{d+2x+2}{d+2x}}&:\text{R}\\[6pt]
  {\displaystyle\tilde{d}^{-1}\frac{1-dq^{2x+2}}{1-dq^{2x}}}
  &:\text{$q$R}\\
  \end{array}\right.,\n
  &\ \ D'(x;\bm{\lambda})=B(x-1;\bm{\lambda})\times\left\{
  \begin{array}{ll}
  {\displaystyle\frac{d+2x-2}{d+2x}}&:\text{R}\\[6pt]
  {\displaystyle q^2\tilde{d}^{-1}\frac{1-dq^{2x-2}}{1-dq^{2x}}}
  &:\text{$q$R}\\
  \end{array}\right..
  \label{B'D'(ii)2}
\end{align}
We introduce the pseudo virtual state polynomial
$\check{\xi}_{\text{v}}(x;\bm{\lambda})$ ($\text{v}\in\mathbb{Z}_{\geq0}$),
\begin{equation}
  (\text{\romannumeral2}):
  \ \ \check{\xi}_{\text{v}}(x;\bm{\lambda})\eqdef
  \check{P}_{\text{v}}\bigl(x-N-1;\mathfrak{t}(\bm{\lambda})\bigr),
  \label{xi(ii)}
\end{equation}
which is the polynomial part of the pseudo virtual state vector,
see Appendix\,\ref{app:pvsv}.
It is a polynomial of degree $\text{v}$ in $\eta(x;\bm{\lambda})$,
$\check{\xi}_{\text{v}}(x;\bm{\lambda})\eqdef
\xi_{\text{v}}\bigl(\eta(x;\bm{\lambda});\bm{\lambda}\bigr)$, because
$\eta\bigl(\mathfrak{t}(x);\mathfrak{t}(\bm{\lambda})\bigr)
=a\eta(x;\bm{\lambda})+b$ ($a,b$ : constants, $a\neq 0$).
Explicitly it is
\begin{equation}
  (\text{\romannumeral2}):
  \ \ \check{\xi}_{\text{v}}(x;\bm{\lambda})
  =\left\{
  \begin{array}{ll}
  {\displaystyle
  {}_4F_3\Bigl(
  \genfrac{}{}{0pt}{}{-\text{v},\,\text{v}+2-\tilde{d},\,1-x-c,
  \,x+a+b-\tilde{d}}
  {1+a-\tilde{d},\,1+b-\tilde{d},\,2-c}\Bigm|1\Bigr)}&:\text{R}\\[10pt]
  {\displaystyle
  {}_4\phi_3\Bigl(
  \genfrac{}{}{0pt}{}{q^{-\text{v}},\,\tilde{d}^{-1}q^{\text{v}+2},
  \,c^{-1}q^{1-x},\,ab\tilde{d}^{-1}q^x}
  {qa\tilde{d}^{-1},\,qb\tilde{d}^{-1},\,q^2c^{-1}}\Bigm|q\,;q\Bigr)}
  &:\text{$q$R}
  \end{array}
  \right..
\end{equation}

By applying the symmetry \eqref{sym0} to this twist (\romannumeral2),
we define a twist (\romannumeral1):
\begin{equation}
  (\text{\romannumeral1}):
  \ \ \mathfrak{t}(x)\eqdef-x-1,
  \ \ \mathfrak{t}(\bm{\lambda})
  \eqdef(2-\lambda_1,2-\lambda_2,2-\lambda_3,2-\lambda_4)
  =2\bm{\delta}-\bm{\lambda},
  \ \ \mathfrak{t}(q)\eqdef q.
  \label{twist(i)}
\end{equation}
By using this twist (\romannumeral1), the functions $B'(x)$ and $D'(x)$
satisfying \eqref{BD=B'D'}--\eqref{B+D=B'+D'} are obtained:
\begin{equation}
  (\text{\romannumeral1}):
  \ \ B'(x;\bm{\lambda})\eqdef D\bigl(-x-1;\mathfrak{t}(\bm{\lambda})\bigr),
  \ \ D'(x;\bm{\lambda})\eqdef B\bigl(-x-1;\mathfrak{t}(\bm{\lambda})\bigr).
  \label{B'D'(i)}
\end{equation}
Since the inversion of the coordinate $x$ ($x\to-x-1$) means the exchange of
the matrices $e^{\partial}\leftrightarrow e^{-\partial}$, the functions
$B(x)$ and $D(x)$ are exchanged in this definition.
Explicit forms of $B'(x;\bm{\lambda})$ and $D'(x;\bm{\lambda})$ are the same
as \eqref{B'D'(ii)2},
\begin{equation}
  B^{\prime\,(\text{\romannumeral1})}(x;\bm{\lambda})
  =B^{\prime\,(\text{\romannumeral2})}(x;\bm{\lambda}),\quad
  D^{\prime\,(\text{\romannumeral1})}(x;\bm{\lambda})
  =D^{\prime\,(\text{\romannumeral2})}(x;\bm{\lambda}),
  \label{B'D'(i)(ii)}
\end{equation}
and $\alpha(\bm{\lambda})$ and $\alpha'(\bm{\lambda})$ are given by
\eqref{alal'(ii)}.
The pseudo virtual state polynomial
$\check{\xi}_{\text{v}}(x;\bm{\lambda})$ ($\text{v}\in\mathbb{Z}_{\geq0}$)
is defined by
\begin{equation}
  (\text{\romannumeral1}):
  \ \ \check{\xi}_{\text{v}}(x;\bm{\lambda})\eqdef
  \check{P}_{\text{v}}\bigl(-x-1;\mathfrak{t}(\bm{\lambda})\bigr),
  \label{xi(i)}
\end{equation}
which should be proportional to the twist (\romannumeral2) case.
In fact, we have
\begin{equation}
  (\text{\romannumeral1}):
  \ \ \check{\xi}_{\text{v}}(x;\bm{\lambda})
  =\left\{
  \begin{array}{ll}
  {\displaystyle
  {}_4F_3\Bigl(
  \genfrac{}{}{0pt}{}{-\text{v},\,\text{v}+2-\tilde{d},\,x+1,\,1-x-d}
  {2-a,\,2-b,\,2-c}\Bigm|1\Bigr)}&:\text{R}\\[10pt]
  {\displaystyle
  {}_4\phi_3\Bigl(
  \genfrac{}{}{0pt}{}{q^{-{\text{v}}},\,\tilde{d}^{-1}q^{\text{v}+2},
  \,q^{x+1},\,d^{-1}q^{1-x}}
  {q^2a^{-1},\,q^2b^{-1},\,q^2c^{-1}}\Bigm|q\,;q\Bigr)}&:\text{$q$R}
  \end{array}
  \right.,
  \label{xi(i)qR}
\end{equation}
and
\begin{equation}
  \check{\xi}^{\text{(\romannumeral2)}}_{\text{v}}(x;\bm{\lambda})
  =\check{\xi}^{\text{(\romannumeral1)}}_{\text{v}}(x;\bm{\lambda})
  \times\left\{
  \begin{array}{ll}
  {\displaystyle
  \frac{(2-a,2-b)_{\text{v}}}{(1+a-\tilde{d},1+b-\tilde{d})_{\text{v}}}}
  &:\text{R}\\[12pt]
  {\displaystyle
  \frac{d^{\text{v}}(q^2a^{-1},q^2b^{-1};q)_{\text{v}}}
  {c^{\text{v}}(qa\tilde{d}^{-1},qb\tilde{d}^{-1};q)_{\text{v}}}}
  &:\text{$q$R} 
  \end{array}
  \right..
  \label{xi(ii)=xi(i)}
\end{equation}

For $q$R case, we can change the parameter $q$.
By applying the symmetry \eqref{sym0t} to the twists
(\romannumeral1)--(\romannumeral2),
we define the twists
($\widetilde{\text{\romannumeral1}}$)--($\widetilde{\text{\romannumeral2}}$)
for $q$R as follows:
\begin{align}
  (\widetilde{\text{\romannumeral1}}):&
  \ \ (\text{\romannumeral1})\text{ with the replacement }
  \mathfrak{t}(q)\eqdef q^{-1},
  \label{twist(i)t}\\
  (\widetilde{\text{\romannumeral2}}):&
  \ \ (\text{\romannumeral2})\text{ with the replacement }
  \mathfrak{t}(q)\eqdef q^{-1},
  \label{twist(ii)t}
\end{align}
which give
\begin{align}
  &\,(\widetilde{\text{\romannumeral1}}):\ B'(x;\bm{\lambda};q)\eqdef
  D\bigl(-x-1;\mathfrak{t}(\bm{\lambda});q^{-1}\bigr),
  \ D'(x;\bm{\lambda};q)\eqdef
  B\bigl(-x-1;\mathfrak{t}(\bm{\lambda});q^{-1}\bigr),
  \label{B'D'(i)t}\\
  &(\widetilde{\text{\romannumeral2}}):\ B'(x;\bm{\lambda};q)\eqdef
  B\bigl(x-N-1;\mathfrak{t}(\bm{\lambda});q^{-1}\bigr),
  \ D'(x;\bm{\lambda};q)\eqdef
  D\bigl(x-N-1;\mathfrak{t}(\bm{\lambda});q^{-1}\bigr),
  \label{B'D'(ii)t}\\
  &(\widetilde{\text{\romannumeral1}}),(\widetilde{\text{\romannumeral2}}):
  \ \alpha(\bm{\lambda})=q,\quad
  \alpha'(\bm{\lambda})=-(1-q)(1-\tilde{d}q^{-1}).
\end{align}
Explicitly $B'(x)$ and $D'(x)$ are
\begin{equation}
  (\widetilde{\text{\romannumeral1}}),(\widetilde{\text{\romannumeral2}}):
  \ \ \alpha(\bm{\lambda})B'(x;\bm{\lambda})=
  \alpha^{(\text{\romannumeral1})}(\bm{\lambda})
  B^{\prime\,(\text{\romannumeral1})}(x;\bm{\lambda}),\quad
  \alpha(\bm{\lambda})D'(x;\bm{\lambda})=
  \alpha^{(\text{\romannumeral1})}(\bm{\lambda})
  D^{\prime\,(\text{\romannumeral1})}(x;\bm{\lambda}).
  \label{B'D'(i)t(ii)t}
\end{equation}
The pseudo virtual state polynomials
$\check{\xi}_{\text{v}}(x;\bm{\lambda})$ ($\text{v}\in\mathbb{Z}_{\geq0}$)
are defined by
\begin{align}
  (\widetilde{\text{\romannumeral1}}):&
  \ \ \check{\xi}_{\text{v}}(x;\bm{\lambda};q)\eqdef
  \check{P}_{\text{v}}\bigl(-x-1;\mathfrak{t}(\bm{\lambda});q^{-1}\bigr),
  \label{xi(i)t}\\
  (\widetilde{\text{\romannumeral2}}):&
  \ \ \check{\xi}_{\text{v}}(x;\bm{\lambda};q)\eqdef
  \check{P}_{\text{v}}\bigl(x-N-1;\mathfrak{t}(\bm{\lambda});q^{-1}\bigr),
  \label{xi(ii)t}
\end{align}
and \eqref{qRq^{-1}} implies
\begin{equation}
  \check{\xi}^{(\widetilde{\text{\romannumeral1}})}_{\text{v}}
  (x;\bm{\lambda};q)
  =\check{\xi}^{\text{(\romannumeral1)}}_{\text{v}}(x;\bm{\lambda};q),\quad
  \check{\xi}^{(\widetilde{\text{\romannumeral2}})}_{\text{v}}
  (x;\bm{\lambda};q)
  =\check{\xi}^{\text{(\romannumeral2)}}_{\text{v}}(x;\bm{\lambda};q).
  \label{xi(i)t=xi(i)}
\end{equation}

The twist operations (\romannumeral1)--(\romannumeral2) and
($\widetilde{\text{\romannumeral1}}$)--($\widetilde{\text{\romannumeral2}}$)
have essentially the same effects for R and $q$R systems, because the
relations \eqref{B'D'(i)(ii)}, \eqref{B'D'(i)t(ii)t}, \eqref{xi(ii)=xi(i)}
and \eqref{xi(i)t=xi(i)} imply that the pseudo virtual vectors \eqref{phitv}
obtained by these twists are same (proportional).
So they lead to the same Casoratian identities \eqref{casoidqR}.
However, they may have different effects for the reduced systems in
\S\,\ref{sec:Casoidred}, which are obtained as appropriate limits of R and
$q$R systems, because the symmetries \eqref{sym0}--\eqref{sym0t} may no longer
hold for the reduced systems.

Since R and $q$R systems are invariant under the exchange
$a\leftrightarrow b$, twists (\romannumeral1)--(\romannumeral2) and
($\widetilde{\text{\romannumeral1}}$)--($\widetilde{\text{\romannumeral2}}$)
can be modified by exchanging
$\mathfrak{t}(\lambda_1)\leftrightarrow\mathfrak{t}(\lambda_2)$.

\subsection{Casoratian identities}
\label{sec:casoidqR}

We will present the Casoratian identities for R and $q$R polynomials.
By using discrete symmetries obtained in \S\,\ref{sec:discsym}, 
the pseudo virtual state vectors are defined, see Appendix\,\ref{app:pvsv}.
Then the original systems can be deformed by multi-step Darboux
transformations in terms of pseudo virtual state vectors.
The original systems can be also deformed by multi-step Darboux
transformations in terms of eigenstate vectors \cite{gos}.
For appropriate choice of the index sets and shift of parameters,
these two deformed systems are found to be equivalent. 
We present such calculation for one-step Darboux transformation in terms of
the pseudo virtual state vector for $q$-Racah case in Appendix\,\ref{app:pvsv}.
However, multi-step Darboux transformations in terms of pseudo virtual state
vectors are rather complicated due to the following two facts:
(a) the pseudo virtual state vectors do not satisfy the Schr\"odinger
equation at both boundaries, (b) the size of the Hamiltonian increases at
each step (namely the $(N+1)\times(N+1)$ matrix becomes the
$(N+M+1)\times(N+M+1)$ matrix after $M$-step).
So we present and prove the Casoratian identities for R and $q$R polynomials
in a `shortcut' way.

We take $M$, $\mathcal{N}$, $\bar{\mathcal{N}}$, $d_j$, $\bar{d}_j$, $e_j$
and $\bar{\bm{\lambda}}$ as \eqref{Dbar}.
Then the Casoratian identities for R and $q$R polynomials are
\begin{align}
  &\quad
  \varphi_M(x-M;\bm{\lambda})^{-1}\,
  \text{W}_{\text{C}}[\check{\xi}_{d_1},\check{\xi}_{d_2},\ldots,
  \check{\xi}_{d_M}](x-M;\bm{\lambda})\n
  &\propto
  \varphi_{\bar{\mathcal{N}}}(x;\bar{\bm{\lambda}})^{-1}\,
  \text{W}_{\text{C}}[\check{P}_{e_1},\check{P}_{e_2},\ldots,
  \check{P}_{e_{\bar{\mathcal{N}}}}](x;\bar{\bm{\lambda}}).
  \label{casoidqR}
\end{align}
Note that the variable $x$ in the first line is shifted by $-M$,
which corresponds to the range of $x$ in the deformed Hamiltonian
$(\mathcal{H}_{d_1\ldots d_M\,x,y})_{-M\leq x,y\leq N}$,
see Appendix\,\ref{app:pvsv}.
Since $\check{P}_n(x;\bm{\lambda})$, $\check{\xi}_{\text{v}}(x;\bm{\lambda})$
and $\varphi_M(x;\bm{\lambda})$ are defined for real $x$, these identities
hold for real $x$.
For $q$R case, we prove \eqref{casoidqR} by translating the Casoratian
identities \eqref{casoidAW} for the Askey-Wilson polynomial.
The identities for R case is easily obtained from $q$R case by taking
$q\to 1$ limit.
The necessary data of the Askey-Wilson polynomial are given in Appendix
\ref{app:AW}.

The $q$-Racah polynomial and the Askey-Wilson polynomial are the `same'
polynomials \cite{kls}.
The replacement rule of this correspondence is
\begin{align}
  &ix^{\text{AW}}=\gamma(x^{\text{$q$R}}+\tfrac12\lambda^{\text{$q$R}}_4),\quad
  \bm{\lambda}^{\text{AW}}=\bm{\lambda}^{\text{$q$R}}
  -\tfrac12\lambda^{\text{$q$R}}_4\bm{\delta}^{\text{$q$R}},\n
  \text{namely}\ \ &
  e^{ix^{\text{AW}}}=q^{x^{\text{$q$R}}}d^{\frac12},\quad
  (a_1,a_2,a_3,a_4)=(ad^{-\frac12},bd^{-\frac12},cd^{-\frac12},d^{\frac12}).
  \label{rule}
\end{align}
Under this replacement rule, we have
\begin{equation}
  \check{P}^{\text{AW}}_n(x^{\text{AW}};\bm{\lambda}^{\text{AW}})
  =d^{-\frac{n}{2}}(a,b,c;q)_n
  \check{P}^{\text{$q$R}}_n(x^{\text{$q$R}};\bm{\lambda}^{\text{$q$R}}),
  \label{PAW=PqR}
\end{equation}
and
\begin{align}
  V(x^{\text{AW}};\bm{\lambda}^{\text{AW}})
  &=-B(x^{\text{$q$R}};\bm{\lambda}^{\text{$q$R}}),
  \ \ V^*(x^{\text{AW}};\bm{\lambda}^{\text{AW}})
  =-D(x^{\text{$q$R}};\bm{\lambda}^{\text{$q$R}}),\n
  \mathcal{E}^{\text{AW}}_n(\bm{\lambda}^{\text{AW}})
  &=\mathcal{E}^{\text{$q$R}}_n(\bm{\lambda}^{\text{$q$R}}),\n
  \eta^{\text{AW}}(x^{\text{AW}})
  &=\tfrac12d^{-\frac12}\bigl(
  \eta^{\text{$q$R}}(x^{\text{$q$R}};\bm{\lambda}^{\text{$q$R}})+1+d\bigr),\\
  \varphi^{\text{AW}}(x^{\text{AW}})
  &=i(dq)^{-\frac12}(1-dq)
  \varphi^{\text{$q$R}}(x^{\text{$q$R}}-\tfrac12;\bm{\lambda}^{\text{$q$R}}).
  \nonumber
\end{align}
The shifts $\bm{\delta}^{\text{AW}}$ and $\bm{\delta}^{\text{$q$R}}$ are
consistent.
The twist $\mathfrak{t}^{\text{AW}}(\bm{\lambda}^{\text{AW}})$ \eqref{twistAW}
gives the twist (\romannumeral1)
$\mathfrak{t}^{\text{$q$R(\romannumeral1)}}(\bm{\lambda}^{\text{$q$R}})$
\eqref{twist(i)}.
In the following we omit the superscript $q$R.
The twisted potential function of AW system $V'$ \eqref{V'},
the pseudo virtual state energy
$\tilde{\mathcal{E}}^{\text{AW}}_{\text{v}}$ \eqref{EtvAW},
the pseudo virtual state polynomial $\check{\xi}_{\text{v}}^{\text{AW}}$
\eqref{xivAW} and
the auxiliary function $\varphi^{\text{AW}}_M$ \eqref{varphiMAW} become
\begin{align}
  &V\bigl(x^{\text{AW}};\mathfrak{t}^{\text{AW}}
  (\bm{\lambda}^{\text{AW}})\bigr)
  =-D\bigl(-x-1;\mathfrak{t}^{(\text{\romannumeral1})}(\bm{\lambda})\bigr)
  =-B^{'(\text{\romannumeral1})}(x;\bm{\lambda}),\n
  &V^*\bigl(x^{\text{AW}};\mathfrak{t}^{\text{AW}}
  (\bm{\lambda}^{\text{AW}})\bigr)
  =-B\bigl(-x-1;\mathfrak{t}^{(\text{\romannumeral1})}(\bm{\lambda})\bigr)
  =-D^{'(\text{\romannumeral1})}(x;\bm{\lambda}),\\
  &\tilde{\mathcal{E}}^{\text{AW}}_{\text{v}}(\bm{\lambda}^{\text{AW}})
  =\tilde{\mathcal{E}}_{\text{v}}(\bm{\lambda})
  =\mathcal{E}_{-\text{v}-1}(\bm{\lambda}),\\
  &\check{\xi}^{\text{AW}}_{\text{v}}(x^{\text{AW}};\bm{\lambda}^{\text{AW}})
  =(qd^{-\frac12})^{\text{v}}(q^2a^{-1},q^2b^{-1},q^2c^{-1};q)_{\text{v}}
  \check{P}_{\text{v}}
  \bigl(-x-1;\mathfrak{t}^{(\text{\romannumeral1})}(\bm{\lambda})\bigr)
  \propto\check{\xi}^{\text{(\romannumeral1)}}_{\text{v}}(x;\bm{\lambda}),
  \label{xiAW=xiqR}\\
  &\varphi^{\text{AW}}_M(x^{\text{AW}})
  \propto\varphi_M(x-\tfrac{M-1}{2};\bm{\lambda}).
  \label{varphiMAW=qR}
\end{align}
The shifted parameters $\bar{\bm{\lambda}}^{\text{AW}}$ and
$\bar{\bm{\lambda}}$ are also consistent.
For functions
$f_j(x^{\text{AW}};\bm{\lambda}^{\text{AW}})=g_j(x;\bm{\lambda})$,
the Casoratian for idQM $\text{W}_{\gamma}$ \eqref{idQM:Wdef} and
that for rdQM $\text{W}_{\text{C}}$ \eqref{rdQM:Wdef} are related by
\begin{equation}
  \text{W}_{\gamma}[f_1,f_2,\ldots,f_n](x^{\text{AW}};\bm{\lambda}^{\text{AW}})
  =i^{\frac12n(n-1)}
  \text{W}_{\text{C}}[g_1,g_2,\ldots,g_n](x-\tfrac{n-1}{2};\bm{\lambda}).
  \label{Wg=WC}
\end{equation}
By using \eqref{PAW=PqR}, \eqref{xiAW=xiqR}--\eqref{varphiMAW=qR} and
\eqref{Wg=WC}, the Casoratian identities for AW polynomial \eqref{casoidAW}
is rewritten as
\begin{align}
  &\quad
  \varphi_M(x-\tfrac{M-1}{2};\bm{\lambda})^{-1}\,
  \text{W}_{\text{C}}[\check{\xi}_{d_1},\check{\xi}_{d_2},\ldots,
  \check{\xi}_{d_M}](x-\tfrac{M-1}{2};\bm{\lambda})\n
  &\propto
  \varphi_{\bar{\mathcal{N}}}(x+\tfrac{M+1}{2};\bar{\bm{\lambda}})^{-1}\,
  \text{W}_{\text{C}}[\check{P}_{e_1},\check{P}_{e_2},\ldots,
  \check{P}_{e_{\bar{\mathcal{N}}}}](x+\tfrac{M+1}{2};\bar{\bm{\lambda}}).
\end{align}
By the replacement $x\to x-\frac{M+1}{2}$, this gives the Casoratian
identities for $q$R polynomial \eqref{casoidqR}.

Although the proportionality constants of \eqref{casoidqR} are not so
important, we present them for the $q$R case.
By explicit calculation (we assume $d_1<d_2<\cdots<d_M$), we have
\begin{align}
  &\quad\varphi_M(x;\bm{\lambda})^{-1}\,
  \text{W}_{\text{C}}[\check{\xi}^{\text{(\romannumeral1)}}_{d_1},
  \check{\xi}^{\text{(\romannumeral1)}}_{d_2},\ldots,
  \check{\xi}^{\text{(\romannumeral1)}}_{d_M}](x;\bm{\lambda})\n
  &=\prod_{j=1}^M(qd^{-1})^{d_j}
  c_{d_j}\bigl(\mathfrak{t}^{\text{(\romannumeral1)}}(\bm{\lambda})\bigr)
  \cdot q^{-\sum_{j=1}^M(j-1)d_j}\prod_{1\leq i<j\leq M}(1-q^{d_j-d_i})
  \cdot q^{\genfrac{(}{)}{0pt}{}{M}{3}}
  \prod_{i=1}^M(1-dq^i)^{M-i}\n
  &\quad\times\Bigl(\text{a monic polynomial of degree $\ell_{\mathcal{D}}$
  in $\eta\bigl(x;\bm{\lambda}+(M-1)\bm{\delta}\bigr)$}\Bigr),\\
  &\quad\varphi_M(x;\bm{\lambda})^{-1}\,
  \text{W}_{\text{C}}[\check{P}_{d_1},\check{P}_{d_2},\ldots,
  \check{P}_{d_M}](x;\bm{\lambda})\n
  &=\prod_{j=1}^Mc_{d_j}(\bm{\lambda})
  \cdot q^{-\sum_{j=1}^M(j-1)d_j}\prod_{1\leq i<j\leq M}(1-q^{d_j-d_i})
  \cdot q^{\genfrac{(}{)}{0pt}{}{M}{3}}
  \prod_{i=1}^M(1-dq^i)^{M-i}\n
  &\quad\times\Bigl(\text{a monic polynomial of degree $\ell_{\mathcal{D}}$
  in $\eta\bigl(x;\bm{\lambda}+(M-1)\bm{\delta}\bigr)$}\Bigr),
\end{align}
where $\ell_{\mathcal{D}}=\sum_{j=1}^Md_j-\frac12M(M-1)$ and
$c_n(\bm{\lambda})$ is the coefficient of the highest degree term
of $P_n(\eta;\bm{\lambda})$,
$\displaystyle c_n(\bm{\lambda})=\frac{(\tilde{d}q^n;q)_n}{(a,b,c;q)_n}$.
Then the proportionality constants (we assume $d_1<d_2<\cdots<d_M$ and
$e_1<e_2<\cdots<e_{\bar{\mathcal{N}}}$) are given by
\begin{align}
  &\quad
  \varphi_M(x-M;\bm{\lambda})^{-1}\,
  \text{W}_{\text{C}}[\check{\xi}^{\text{(\romannumeral1)}}_{d_1},
  \check{\xi}^{\text{(\romannumeral1)}}_{d_2},\ldots,
  \check{\xi}^{\text{(\romannumeral1)}}_{d_M}](x-M;\bm{\lambda})\n
  &=A\times
  \varphi_{\bar{\mathcal{N}}}(x;\bar{\bm{\lambda}})^{-1}\,
  \text{W}_{\text{C}}[\check{P}_{e_1},\check{P}_{e_2},\ldots,
  \check{P}_{e_{\bar{\mathcal{N}}}}](x;\bar{\bm{\lambda}}),\\[4pt]
  &A=\frac{\prod_{j=1}^M
  c_{d_j}\bigl(\mathfrak{t}^{\text{(\romannumeral1)}}(\bm{\lambda})\bigr)}
  {\prod_{j=1}^{\bar{\mathcal{N}}}c_{e_j}(\bar{\bm{\lambda}})}
  \frac{\prod_{1\leq i<j\leq M}(1-q^{d_j-d_i})}
  {\prod_{1\leq i<j\leq\bar{\mathcal{N}}}(1-q^{e_j-e_i})}
  \frac{\prod_{i=1}^M(1-dq^i)^{M-i}}
  {\prod_{i=1}^{\bar{\mathcal{N}}}
  (1-dq^{-\bar{\mathcal{N}}-M+i})^{\bar{\mathcal{N}}-i}}\n
  &\qquad\times
  d^{-\sum_{j=1}^Md_j}
  q^{\sum_{j=1}^M(M+1-j)d_j+\sum_{j=1}^{\bar{\mathcal{N}}}je_j
  -\frac16M(M-1)(2M-1)
  -\frac16(\bar{\mathcal{N}}-1)\bar{\mathcal{N}}(\bar{\mathcal{N}}+1)}.
\end{align}
Here we have used
$\sum_{j=1}^{\bar{\mathcal{N}}}e_j=\sum_{j=1}^Md_j
+\frac12\mathcal{N}(\mathcal{N}+1)-\mathcal{N}M$,
$\ell_{\bar{\mathcal{D}}}=\ell_{\mathcal{D}}$ and
$\eta\bigl(x;\bar{\bm{\lambda}}+(\bar{\mathcal{N}}-1)\bm{\delta}\bigr)
=q^{-M}\eta\bigl(x-M;\bm{\lambda}+(M-1)\bm{\delta}\bigr)
+(q^{-M}-1)(1-dq^{-1})$.

\section{Casoratian Identities for the Reduced Case Polynomials}
\label{sec:Casoidred}

It is well known that the other members of the Askey scheme polynomials
of a discrete variable can be obtained by reductions from the ($q$-)Racah
polynomials \cite{kls}.
Not only the polynomials themselves but also the Hamiltonians are reduced
in appropriate limits (Overall rescalings may be needed).
Some of the twist operations $\mathfrak{t}$ (\romannumeral1)--(\romannumeral2),
($\widetilde{\text{\romannumeral1}}$)--($\widetilde{\text{\romannumeral2}}$)
of R and $q$R systems are inherited to the reduced systems.
The twist operation $\mathfrak{t}$ is \eqref{twist} and the $q$-part should
be ignored for non $q$-polynomials.
The twists (\romannumeral1)--(\romannumeral2) and
($\widetilde{\text{\romannumeral1}}$)--($\widetilde{\text{\romannumeral2}}$)
act on $x$ and $q$ as
\begin{alignat}{2}
  (\text{\romannumeral1}):&
  \ \ \mathfrak{t}(x)\eqdef-x-1,
  &\ \ \mathfrak{t}(q)&\eqdef q,\\
  (\text{\romannumeral2}):&
  \ \ \mathfrak{t}(x)\eqdef x-N-1,
  &\ \ \mathfrak{t}(q)&\eqdef q,\\
  (\widetilde{\text{\romannumeral1}}):&
  \ \ \mathfrak{t}(x)\eqdef-x-1,
  &\ \ \mathfrak{t}(q)&\eqdef q^{-1},\\
  (\widetilde{\text{\romannumeral2}}):&
  \ \ \mathfrak{t}(x)\eqdef x-N-1,
  &\ \ \mathfrak{t}(q)&\eqdef q^{-1},
\end{alignat}
and $\mathfrak{t}(\bm{\lambda})$ will be given for each polynomial.
For non $q$-polynomials, the twists
($\widetilde{\text{\romannumeral1}}$)--($\widetilde{\text{\romannumeral2}}$)
are irrelevant. For infinite systems, the twists (\romannumeral1) and
($\widetilde{\text{\romannumeral1}}$) should be applied.
By using the twist operation, the potential functions $B'(x;\bm{\lambda})$
and $D'(x;\bm{\lambda})$ are defined by \eqref{B'D'(i)}, \eqref{B'D'(ii)}
and \eqref{B'D'(i)t}--\eqref{B'D'(ii)t},
and the pseudo virtual state polynomials
$\check{\xi}_{\text{v}}(x;\bm{\lambda})$ are defined by
\eqref{xi(i)}, \eqref{xi(ii)} and \eqref{xi(i)t}--\eqref{xi(ii)t},
which are polynomials of degree $\text{v}$ in
$\eta(x;\bm{\lambda})$, $\check{\xi}_{\text{v}}(x;\bm{\lambda})\eqdef
\xi_{\text{v}}\bigl(\eta(x;\bm{\lambda});\bm{\lambda}\bigr)$
(or $\eta(x;\bm{\lambda};q)$, $\check{\xi}_{\text{v}}(x;\bm{\lambda};q)\eqdef
\xi_{\text{v}}\bigl(\eta(x;\bm{\lambda};q);\bm{\lambda};q\bigr)$).
The pseudo virtual state energies
$\tilde{\mathcal{E}}_{\text{v}}(\bm{\lambda})$ are defined in
\eqref{Etv(i)}--\eqref{Etv(i)t} and they satisfy \eqref{Etv=E-v-1}.
Then the Casoratian identities for these reduced case polynomials have the
same form as \eqref{casoidqR},
\begin{align}
  &\quad
  \varphi_M(x-M;\bm{\lambda})^{-1}\,
  \text{W}_{\text{C}}[\check{\xi}_{d_1},\check{\xi}_{d_2},\ldots,
  \check{\xi}_{d_M}](x-M;\bm{\lambda})\n
  &\propto
  \varphi_{\bar{\mathcal{N}}}(x;\bar{\bm{\lambda}})^{-1}\,
  \text{W}_{\text{C}}[\check{P}_{e_1},\check{P}_{e_2},\ldots,
  \check{P}_{e_{\bar{\mathcal{N}}}}](x;\bar{\bm{\lambda}}),
  \label{casoidred}
\end{align}
with the notation \eqref{Dbar}.

The fundamental data for the reduced case polynomials are listed in Appendix
\ref{app:finite}--\ref{app:semiinfinite}.
In the following we present twist operations and explicit forms of the pseudo
virtual state polynomials.
For the ($q$-)Hahn, dual ($q$-)Hahn, ($q$-)Krawtchouk and dual
$q$-Krawtchouk cases, we have two twist operations. Two pseudo virtual
state polynomial $\check{\xi}_{\text{v}}(x;\bm{\lambda})$ obtained by
these twists are proportional and two pairs of potential functions
$(B'(x;\bm{\lambda}),D'(x;\bm{\lambda}))$ are also proportional.
Therefore the pseudo virtual state vectors obtained by these twists are
proportional and the corresponding Casoratian identities are identical.

Some formulas in Appendix\,\ref{app:pvsv} are written under the condition
$\alpha(\bm{\lambda})>0$.
For $\alpha(\bm{\lambda})<0$, slight modifications are needed.
For example, \eqref{B'>0}:
$B'(x;\bm{\lambda})\to\alpha(\bm{\lambda})B'(x;\bm{\lambda})$
and $D'(x;\bm{\lambda})\to\alpha(\bm{\lambda})D'(x;\bm{\lambda})$,
\eqref{phit0}: $\tilde{\phi}_0(x;\bm{\lambda})\to\tilde{\phi}_0(x;\bm{\lambda})
\times\bigl(\text{sgn}\,\alpha(\bm{\lambda})\bigr)^{-x}$, etc.  

\subsection{Finite cases}
\label{sec:finite}

\subsubsection{Hahn (Ha)}
\label{sec:Ha}

We have two twist operations:
\begin{align}
  (\text{\romannumeral1}):&
  \ \ \mathfrak{t}(\bm{\lambda})\eqdef(2-\lambda_1,2-\lambda_2,-2-\lambda_3)
  =2\bm{\delta}-\bm{\lambda},\\
  (\text{\romannumeral2}):&
  \ \ \mathfrak{t}(\bm{\lambda})\eqdef(2-\lambda_2,2-\lambda_1,-2-\lambda_3).
\end{align}
The explicit forms of the pseudo virtual state polynomials are
\begin{align}
  (\text{\romannumeral1}):&
  \ \ \check{\xi}_{\text{v}}(x;\bm{\lambda})
  ={}_3F_2\Bigl(
  \genfrac{}{}{0pt}{}{-\text{v},\,\text{v}+3-a-b,\,x+1}{2-a,\,N+2}
  \Bigm|1\Bigr),\\
  (\text{\romannumeral2}):&
  \ \ \check{\xi}_{\text{v}}(x;\bm{\lambda})
  ={}_3F_2\Bigl(
  \genfrac{}{}{0pt}{}{-\text{v},\,\text{v}+3-a-b,\,N+1-x}{2-b,\,N+2}
  \Bigm|1\Bigr),
\end{align}
which are proportional,
\begin{equation}
  \check{\xi}^{(\text{\romannumeral2})}_{\text{v}}(x;\bm{\lambda})
  =\check{\xi}^{(\text{\romannumeral1})}_{\text{v}}(x;\bm{\lambda})
  \times\frac{(2-a)_{\text{v}}}{(b-\text{v}-1)_{\text{v}}},\quad
  \genfrac{(}{)}{0pt}{}{B^{\prime\,(\text{\romannumeral2})}(x;\bm{\lambda})}
  {D^{\prime\,(\text{\romannumeral2})}(x;\bm{\lambda})}
  =\genfrac{(}{)}{0pt}{}{B^{\prime\,(\text{\romannumeral1})}(x;\bm{\lambda})}
  {D^{\prime\,(\text{\romannumeral1})}(x;\bm{\lambda})},
\end{equation}
and $\alpha^{(\text{\romannumeral1})}(\bm{\lambda})
=\alpha^{(\text{\romannumeral2})}(\bm{\lambda})=1$.

\subsubsection{dual Hahn (dHa)}
\label{sec:dHa}

We have two twist operations:
\begin{align}
  (\text{\romannumeral1}):&
  \ \ \mathfrak{t}(\bm{\lambda})\eqdef(2-\lambda_1,2-\lambda_2,-2-\lambda_3),\\
  (\text{\romannumeral2}):&
  \ \ \mathfrak{t}(\bm{\lambda})\eqdef
  (1+\lambda_2+\lambda_3,1+\lambda_1+\lambda_3,-2-\lambda_3).
\end{align}
The explicit forms of the pseudo virtual state polynomials are
\begin{align}
  (\text{\romannumeral1}):&
  \ \ \check{\xi}_{\text{v}}(x;\bm{\lambda})
  ={}_3F_2\Bigl(
  \genfrac{}{}{0pt}{}{-\text{v},\,2-x-a-b,\,x+1}{2-a,\,N+2}
  \Bigm|1\Bigr),\\
  (\text{\romannumeral2}):&
  \ \ \check{\xi}_{\text{v}}(x;\bm{\lambda})
  ={}_3F_2\Bigl(
  \genfrac{}{}{0pt}{}{-\text{v},\,x+a+b+N,\,N+1-x}{b+N+1,\,N+2}
  \Bigm|1\Bigr),
\end{align}
which are proportional,
\begin{equation}
  \check{\xi}^{(\text{\romannumeral2})}_{\text{v}}(x;\bm{\lambda})
  =\check{\xi}^{(\text{\romannumeral1})}_{\text{v}}(x;\bm{\lambda})
  \times\frac{(2-a)_{\text{v}}}{(b+N+1)_{\text{v}}},\quad
  \genfrac{(}{)}{0pt}{}{B^{\prime\,(\text{\romannumeral2})}(x;\bm{\lambda})}
  {D^{\prime\,(\text{\romannumeral2})}(x;\bm{\lambda})}
  =\genfrac{(}{)}{0pt}{}{B^{\prime\,(\text{\romannumeral1})}(x;\bm{\lambda})}
  {D^{\prime\,(\text{\romannumeral1})}(x;\bm{\lambda})},
\end{equation}
and $\alpha^{(\text{\romannumeral1})}(\bm{\lambda})
=\alpha^{(\text{\romannumeral2})}(\bm{\lambda})=-1$.

\subsubsection{Krawtchouk (K)}
\label{sec:K}

We have two twist operations:
\begin{align}
  (\text{\romannumeral1}):&
  \ \ \mathfrak{t}(\bm{\lambda})\eqdef(\lambda_1,-2-\lambda_2),\\
  (\text{\romannumeral2}):&
  \ \ \mathfrak{t}(\bm{\lambda})\eqdef
  (1-\lambda_1,-2-\lambda_2).
\end{align}
The explicit forms of the pseudo virtual state polynomials are
\begin{align}
  (\text{\romannumeral1}):&
  \ \ \check{\xi}_{\text{v}}(x;\bm{\lambda})
  ={}_2F_1\Bigl(
  \genfrac{}{}{0pt}{}{-\text{v},\,x+1}{N+2}
  \Bigm|p^{-1}\Bigr),\\
  (\text{\romannumeral2}):&
  \ \ \check{\xi}_{\text{v}}(x;\bm{\lambda})
  ={}_2F_1\Bigl(
  \genfrac{}{}{0pt}{}{-\text{v},\,N+1-x}{N+2}
  \Bigm|(1-p)^{-1}\Bigr),
\end{align}
which are proportional,
\begin{equation}
  \check{\xi}^{(\text{\romannumeral2})}_{\text{v}}(x;\bm{\lambda})
  =\check{\xi}^{(\text{\romannumeral1})}_{\text{v}}(x;\bm{\lambda})
  \times(1-p^{-1})^{-\text{v}},\quad
  \genfrac{(}{)}{0pt}{}{B^{\prime\,(\text{\romannumeral2})}(x;\bm{\lambda})}
  {D^{\prime\,(\text{\romannumeral2})}(x;\bm{\lambda})}
  =\genfrac{(}{)}{0pt}{}{B^{\prime\,(\text{\romannumeral1})}(x;\bm{\lambda})}
  {D^{\prime\,(\text{\romannumeral1})}(x;\bm{\lambda})},
\end{equation}
and $\alpha^{(\text{\romannumeral1})}(\bm{\lambda})
=\alpha^{(\text{\romannumeral2})}(\bm{\lambda})=-1$.

\subsubsection{$q$-Hahn ($q$Ha)}
\label{sec:qHa}

We have two twist operations:
\begin{align}
  (\widetilde{\text{\romannumeral1}}):&
  \ \ \mathfrak{t}(\bm{\lambda})\eqdef(2-\lambda_1,2-\lambda_2,-2-\lambda_3)
  =2\bm{\delta}-\bm{\lambda},\\
  (\text{\romannumeral2}):&
  \ \ \mathfrak{t}(\bm{\lambda})\eqdef
  (2-\lambda_2,2-\lambda_1,-2-\lambda_3).
\end{align}
The explicit forms of the pseudo virtual state polynomials are
\begin{align}
  (\widetilde{\text{\romannumeral1}}):&
  \ \ \check{\xi}_{\text{v}}(x;\bm{\lambda};q)
  ={}_3\phi_2\Bigl(
  \genfrac{}{}{0pt}{}{q^{-\text{v}},\,a^{-1}b^{-1}q^{\text{v}+3},\,q^{x+1}}
  {q^2a^{-1},\,q^{N+2}}\Bigm|q\,;bq^{N-x}\Bigr),\\[2pt]
  (\text{\romannumeral2}):&
  \ \ \check{\xi}_{\text{v}}(x;\bm{\lambda})
  ={}_3\phi_2\Bigl(
  \genfrac{}{}{0pt}{}{q^{-\text{v}},\,a^{-1}b^{-1}q^{\text{v}+3},\,q^{N+1-x}}
  {q^2b^{-1},\,q^{N+2}}\Bigm|q\,;q\Bigr),
\end{align}
which are proportional,
\begin{equation}
  \check{\xi}^{(\text{\romannumeral2})}_{\text{v}}(x;\bm{\lambda})
  =\check{\xi}^{(\widetilde{\text{\romannumeral1}})}_{\text{v}}(x;\bm{\lambda})
  \times\frac{(q^2a^{-1};q)_{\text{v}}}{(bq^{-\text{v}-1};q)_{\text{v}}},
  \ \ \alpha^{(\text{\romannumeral2})}(\bm{\lambda})
  \genfrac{(}{)}{0pt}{}{B^{\prime\,(\text{\romannumeral2})}(x;\bm{\lambda})}
  {D^{\prime\,(\text{\romannumeral2})}(x;\bm{\lambda})}
  =\alpha^{(\widetilde{\text{\romannumeral1}})}(\bm{\lambda})
  \genfrac{(}{)}{0pt}{}{B^{\prime\,(\widetilde{\text{\romannumeral1}})}
  (x;\bm{\lambda})}
  {D^{\prime\,(\widetilde{\text{\romannumeral1}})}(x;\bm{\lambda})},
\end{equation}
and $\alpha^{(\widetilde{\text{\romannumeral1}})}(\bm{\lambda})=q$,
$\alpha^{(\text{\romannumeral2})}(\bm{\lambda})=abq^{-2}$.

\subsubsection{dual $q$-Hahn (d$q$Ha)}
\label{sec:dqHa}

We have two twist operations:
\begin{align}
  (\widetilde{\text{\romannumeral1}}):&
  \ \ \mathfrak{t}(\bm{\lambda})\eqdef(2-\lambda_1,2-\lambda_2,-2-\lambda_3),\\
  (\widetilde{\text{\romannumeral2}}):&
  \ \ \mathfrak{t}(\bm{\lambda})\eqdef
  (1+\lambda_2+\lambda_3,1+\lambda_1+\lambda_3,-2-\lambda_3).
\end{align}
The explicit forms of the pseudo virtual state polynomials are
\begin{align}
  (\widetilde{\text{\romannumeral1}}):&
  \ \ \check{\xi}_{\text{v}}(x;\bm{\lambda};q)
  ={}_3\phi_2\Bigl(
  \genfrac{}{}{0pt}{}{q^{-\text{v}},\,a^{-1}b^{-1}q^{2-x},\,q^{x+1}}
  {q^2a^{-1},\,q^{N+2}}\Bigm|q\,;bq^{\text{v}+N+1}\Bigr),\\[2pt]
  (\widetilde{\text{\romannumeral2}}):&
  \ \ \check{\xi}_{\text{v}}(x;\bm{\lambda};q)
  ={}_3\phi_2\Bigl(
  \genfrac{}{}{0pt}{}{q^{-\text{v}},\,abq^{x+N},\,q^{N+1-x}}
  {bq^{N+1},\,q^{N+2}}\Bigm|q\,;a^{-1}q^{\text{v}+2}\Bigr),
\end{align}
which are proportional,
\begin{equation}
  \check{\xi}^{(\widetilde{\text{\romannumeral2}})}_{\text{v}}(x;\bm{\lambda})
  =\check{\xi}^{(\widetilde{\text{\romannumeral1}})}_{\text{v}}(x;\bm{\lambda})
  \times\frac{(q^2a^{-1};q)_{\text{v}}}{(bq^{N+1};q)_{\text{v}}},\quad
  \genfrac{(}{)}{0pt}{}{B^{\prime\,(\widetilde{\text{\romannumeral2}})}
  (x;\bm{\lambda})}
  {D^{\prime\,(\widetilde{\text{\romannumeral2}})}(x;\bm{\lambda})}
  =\genfrac{(}{)}{0pt}{}{B^{\prime\,(\widetilde{\text{\romannumeral1}})}
  (x;\bm{\lambda})}
  {D^{\prime\,(\widetilde{\text{\romannumeral1}})}(x;\bm{\lambda})},
\end{equation}
and $\alpha^{(\widetilde{\text{\romannumeral1}})}(\bm{\lambda})=
\alpha^{(\widetilde{\text{\romannumeral2}})}(\bm{\lambda})=q$.

\subsubsection{quantum $q$-Krawtchouk (q$q$K)}
\label{sec:qqK}

We have one twist operation:
\begin{equation}
  (\widetilde{\text{\romannumeral1}}):
  \ \ \mathfrak{t}(\bm{\lambda})\eqdef(-\lambda_1,-2-\lambda_2).
\end{equation}
The explicit form of the pseudo virtual state polynomials is
\begin{equation}
  (\widetilde{\text{\romannumeral1}}):
  \ \ \check{\xi}_{\text{v}}(x;\bm{\lambda};q)
  ={}_2\phi_1\Bigl(
  \genfrac{}{}{0pt}{}{q^{-\text{v}},\,q^{x+1}}
  {q^{N+2}}\Bigm|q\,;pq^{N+1-x}\Bigr).
\end{equation}

\subsubsection{$q$-Krawtchouk ($q$K)}
\label{sec:qK}

We have two twist operations with different $\mathfrak{t}(q)$'s:
\begin{align}
  (\widetilde{\text{\romannumeral1}}):&
  \ \ \mathfrak{t}(\bm{\lambda})\eqdef(2-\lambda_1,-2-\lambda_2),\quad
  \mathfrak{t}(q)=q^{-1},\\
  (\text{\romannumeral2}):&
  \ \ \mathfrak{t}(\bm{\lambda})\eqdef
  (2-\lambda_1,-2-\lambda_2),\quad\mathfrak{t}(q)=q.
\end{align}
The explicit forms of the pseudo virtual state polynomials are
\begin{align}
  (\widetilde{\text{\romannumeral1}}):&
  \ \ \check{\xi}_{\text{v}}(x;\bm{\lambda};q)
  ={}_3\phi_1\Bigl(
  \genfrac{}{}{0pt}{}{q^{-\text{v}},\,-p^{-1}q^{\text{v}+2},\,q^{x+1}}
  {q^{N+2}}\Bigm|q\,;-pq^{N-x-1}\Bigr),\\[2pt]
  (\text{\romannumeral2}):&
  \ \ \check{\xi}_{\text{v}}(x;\bm{\lambda})
  ={}_3\phi_2\Bigl(
  \genfrac{}{}{0pt}{}{q^{-\text{v}},\,-p^{-1}q^{\text{v}+2},\,q^{N+1-x}}
  {q^{N+2},\,0}\Bigm|q\,;q\Bigr),
\end{align}
which are proportional,
\begin{equation}
  \check{\xi}^{(\text{\romannumeral2})}_{\text{v}}(x;\bm{\lambda})
  =\check{\xi}^{(\widetilde{\text{\romannumeral1}})}_{\text{v}}(x;\bm{\lambda})
  \times(-p)^{-\text{v}}q^{\text{v}(\text{v}+2)},
  \ \ \alpha^{(\text{\romannumeral2})}(\bm{\lambda})
  \genfrac{(}{)}{0pt}{}{B^{\prime\,(\text{\romannumeral2})}(x;\bm{\lambda})}
  {D^{\prime\,(\text{\romannumeral2})}(x;\bm{\lambda})}
  =\alpha^{(\widetilde{\text{\romannumeral1}})}(\bm{\lambda})
  \genfrac{(}{)}{0pt}{}{B^{\prime\,(\widetilde{\text{\romannumeral1}})}
  (x;\bm{\lambda})}
  {D^{\prime\,(\widetilde{\text{\romannumeral1}})}(x;\bm{\lambda})},
\end{equation}
and $\alpha^{(\widetilde{\text{\romannumeral1}})}(\bm{\lambda})=q$,
$\alpha^{(\text{\romannumeral2})}(\bm{\lambda})=-pq^{-1}$.

\subsubsection{dual $q$-Krawtchouk (d$q$K)}
\label{sec:dqK}

We have two twist operations:
\begin{align}
  (\widetilde{\text{\romannumeral1}}):&
  \ \ \mathfrak{t}(\bm{\lambda})\eqdef(-\lambda_1,-2-\lambda_2),\\
  (\widetilde{\text{\romannumeral2}}):&
  \ \ \mathfrak{t}(\bm{\lambda})\eqdef
  (\lambda_1,-2-\lambda_2).
\end{align}
The explicit forms of the pseudo virtual state polynomials are
\begin{align}
  (\widetilde{\text{\romannumeral1}}):&
  \ \ \check{\xi}_{\text{v}}(x;\bm{\lambda};q)
  ={}_3\phi_1\Bigl(
  \genfrac{}{}{0pt}{}{q^{-\text{v}},c^{-1}q^{N+1-x},\,q^{x+1}}
  {q^{N+2}}\Bigm|q\,;cq^{\text{v}}\Bigr),\\[2pt]
  (\widetilde{\text{\romannumeral2}}):&
  \ \ \check{\xi}_{\text{v}}(x;\bm{\lambda};q)
  ={}_3\phi_1\Bigl(
  \genfrac{}{}{0pt}{}{q^{-\text{v}},\,cq^{x+1},\,q^{N+1-x}}
  {q^{N+2}}\Bigm|q\,;c^{-1}q^{\text{v}}\Bigr),
\end{align}
which are proportional,
\begin{equation}
  \check{\xi}^{(\widetilde{\text{\romannumeral2}})}_{\text{v}}(x;\bm{\lambda})
  =\check{\xi}^{(\widetilde{\text{\romannumeral1}})}_{\text{v}}(x;\bm{\lambda})
  \times c^{-\text{v}},\quad
  \genfrac{(}{)}{0pt}{}{B^{\prime\,(\widetilde{\text{\romannumeral2}})}
  (x;\bm{\lambda})}
  {D^{\prime\,(\widetilde{\text{\romannumeral2}})}(x;\bm{\lambda})}
  =\genfrac{(}{)}{0pt}{}{B^{\prime\,(\widetilde{\text{\romannumeral1}})}
  (x;\bm{\lambda})}
  {D^{\prime\,(\widetilde{\text{\romannumeral1}})}(x;\bm{\lambda})},
\end{equation}
and $\alpha^{(\widetilde{\text{\romannumeral1}})}(\bm{\lambda})=
\alpha^{(\widetilde{\text{\romannumeral2}})}(\bm{\lambda})=q$.

\subsubsection{affine $q$-Krawtchouk (a$q$K)}
\label{sec:aqK}

We have one twist operation:
\begin{equation}
  (\widetilde{\text{\romannumeral1}}):
  \ \ \mathfrak{t}(\bm{\lambda})\eqdef(-\lambda_1,-2-\lambda_2).
\end{equation}
The explicit form of the pseudo virtual state polynomials is
\begin{equation}
  (\widetilde{\text{\romannumeral1}}):
  \ \ \check{\xi}_{\text{v}}(x;\bm{\lambda};q)
  ={}_2\phi_2\Bigl(
  \genfrac{}{}{0pt}{}{q^{-\text{v}},\,q^{x+1}}
  {p^{-1}q,\,q^{N+2}}\Bigm|q\,;p^{-1}q^{\text{v}+N+2-x}\Bigr).
\end{equation}

\subsection{Semi-infinite cases}
\label{sec:semiinfinite}

\subsubsection{Meixner (M)}
\label{sec:M}

We have one twist operation:
\begin{equation}
  (\text{\romannumeral1}):
  \ \ \mathfrak{t}(\bm{\lambda})\eqdef(2-\lambda_1,\lambda_2).
\end{equation}
The explicit form of the pseudo virtual state polynomials is
\begin{equation}
  (\text{\romannumeral1}):
  \ \ \check{\xi}_{\text{v}}(x;\bm{\lambda})
  ={}_2F_1\Bigl(
  \genfrac{}{}{0pt}{}{-\text{v},\,x+1}{2-\beta}\Bigm|1-c^{-1}\Bigr).
\end{equation}

\subsubsection{Charlier (C)}
\label{sec:C}

We have one twist operation:
\begin{equation}
  (\text{\romannumeral1}):
  \ \ \mathfrak{t}(\bm{\lambda})\eqdef-\lambda_1.
\end{equation}
The explicit form of the pseudo virtual state polynomials is
\begin{equation}
  (\text{\romannumeral1}):
  \ \ \check{\xi}_{\text{v}}(x;\bm{\lambda})
  ={}_2F_0\Bigl(
  \genfrac{}{}{0pt}{}{-\text{v},\,x+1}{-}\Bigm|a^{-1}\Bigr).
\end{equation}

\subsubsection{little $q$-Jacobi (l$q$J)}
\label{sec:lqJ}

We have one twist operation:
\begin{equation}
  (\widetilde{\text{\romannumeral1}}):
  \ \ \mathfrak{t}(\bm{\lambda})\eqdef(-\lambda_1,-\lambda_2).
\end{equation}
The explicit forms of the pseudo virtual state polynomials are
\begin{align}
  (\widetilde{\text{\romannumeral1}}):
  \ \ \check{\xi}_{\text{v}}(x;\bm{\lambda};q)
  &=(-b)^{-\text{v}}q^{\frac12\text{v}(\text{v}+1)}
  \frac{(a^{-1}q;q)_{\text{v}}}{(b^{-1}q;q)_{\text{v}}}\,
  {}_2\phi_1\Bigl(
  \genfrac{}{}{0pt}{}{q^{-\text{v}},\,a^{-1}b^{-1}q^{\text{v}+1}}
  {a^{-1}q}\Bigm|q\,;bq^{x+1}\Bigr)\n
  &={}_3\phi_2\Bigl(
  \genfrac{}{}{0pt}{}{q^{-\text{v}},\,a^{-1}b^{-1}q^{\text{v}+1},\,q^{x+1}}
  {b^{-1}q,\,0}\Bigm|q\,;q\Bigr).
\end{align}

\subsubsection{$q$-Meixner ($q$M)}
\label{sec:qM}

We have one twist operation:
\begin{equation}
  (\widetilde{\text{\romannumeral1}}):
  \ \ \mathfrak{t}(\bm{\lambda})\eqdef(-\lambda_1,-\lambda_2).
\end{equation}
The explicit form of the pseudo virtual state polynomials is
\begin{equation}
  (\widetilde{\text{\romannumeral1}}):
  \ \ \check{\xi}_{\text{v}}(x;\bm{\lambda};q)
  ={}_2\phi_1\Bigl(
  \genfrac{}{}{0pt}{}{q^{-\text{v}},\,q^{x+1}}
  {b^{-1}q}\Bigm|q\,;-b^{-1}c^{-1}q^{-x}\Bigr).
\end{equation}

\subsubsection{little $q$-Laguerre/Wall (l$q$L)}
\label{sec:lqL}

We have one twist operation:
\begin{equation}
  (\widetilde{\text{\romannumeral1}}):
  \ \ \mathfrak{t}(\bm{\lambda})\eqdef-\lambda_1.
\end{equation}
The explicit forms of the pseudo virtual state polynomials are
\begin{align}
  (\widetilde{\text{\romannumeral1}}):
  \ \ \check{\xi}_{\text{v}}(x;\bm{\lambda};q)
  &={}_2\phi_1\Bigl(
  \genfrac{}{}{0pt}{}{q^{-\text{v}},\,q^{x+1}}
  {0}\Bigm|q\,;a^{-1}q^{\text{v}+1}\Bigr)\n
  &=(-a)^{-\text{v}}q^{\frac12\text{v}(\text{v}+1)}
  (aq^{-\text{v}};q)_{\text{v}}\,
  {}_1\phi_1\Bigl(
  \genfrac{}{}{0pt}{}{q^{-\text{v}}}
  {a^{-1}q}\Bigm|q\,;a^{-1}q^{\text{v}+x+2}\Bigr).
\end{align}

\subsubsection{Al-Salam-Carlitz \Romannumeral{2} (ASCII)}
\label{sec:ASCII}

We have one twist operation:
\begin{equation}
  (\widetilde{\text{\romannumeral1}}):
  \ \ \mathfrak{t}(\bm{\lambda})\eqdef-\lambda_1.
\end{equation}
The explicit form of the pseudo virtual state polynomials is
\begin{equation}
  (\widetilde{\text{\romannumeral1}}):
  \ \ \check{\xi}_{\text{v}}(x;\bm{\lambda};q)
  ={}_2\phi_1\Bigl(
  \genfrac{}{}{0pt}{}{q^{-\text{v}},\,q^{x+1}}
  {0}\Bigm|q\,;a^{-1}q^{-x}\Bigr).
\end{equation}

\subsubsection{$q$-Bessel ($q$B) (alternative $q$-Charlier)}
\label{sec:aqC}

We have one twist operation:
\begin{equation}
  (\widetilde{\text{\romannumeral1}}):
  \ \ \mathfrak{t}(\bm{\lambda})\eqdef 2-\lambda_1.
\end{equation}
The explicit forms of the pseudo virtual state polynomials are
\begin{align}
  (\widetilde{\text{\romannumeral1}}):
  \ \ \check{\xi}_{\text{v}}(x;\bm{\lambda};q)
  &=q^{\text{v}(x+1)}
  {}_2\phi_0\Bigl(\genfrac{}{}{0pt}{}{q^{-\text{v}},\,q^{x+1}}
  {-}\Bigm|q\,;-a^{-1}q^{2\text{v}+1-x}\Bigr)\n
  &=(-a)^{-\text{v}}q^{\text{v}(\text{v}+2)}
  {}_2\phi_0\Bigl(
  \genfrac{}{}{0pt}{}{q^{-\text{v}},\,-a^{-1}q^{\text{v}+2}}
  {-}\Bigm|q\,;-aq^{x-1}\Bigr)\\
  &={}_3\phi_2\Bigl(
  \genfrac{}{}{0pt}{}{q^{-\text{v}},\,-a^{-1}q^{\text{v}+2},\,q^{x+1}}
  {0,0}\Bigm|q\,;q\Bigr).\nonumber
\end{align}

\subsubsection{$q$-Charlier ($q$C)}
\label{sec:qC}

We have one twist operation:
\begin{equation}
  (\widetilde{\text{\romannumeral1}}):
  \ \ \mathfrak{t}(\bm{\lambda})\eqdef-\lambda_1.
\end{equation}
The explicit form of the pseudo virtual state polynomials is
\begin{equation}
  (\widetilde{\text{\romannumeral1}}):
  \ \ \check{\xi}_{\text{v}}(x;\bm{\lambda};q)
  ={}_2\phi_0\Bigl(
  \genfrac{}{}{0pt}{}{q^{-\text{v}},\,q^{x+1}}
  {-}\Bigm|q\,;-a^{-1}q^{-x-1}\Bigr).
\end{equation}

\section{Summary and Comments}
\label{sec:summary}

In addition to the Wronskian identities for the Hermite, Laguerre and
Jacobi polynomials in oQM \cite{os29} and the Casoratian identities for
the Askey-Wilson polynomial and its reduced form polynomials in idQM
\cite{os30}, infinitely many Casoratian identities for the $q$-Racah
polynomial and its reduced form polynomials in rdQM are obtained.
The pseudo virtual state polynomials are defined by using discrete
symmetries of the original systems.
The derivation of the Casoratian identities in this paper is a `shortcut' way.
The pseudo virtual state vectors and the one-step Darboux transformation
in terms of it for $q$-Racah case are discussed in Appendix\,\ref{app:pvsv}.
We will report on the multi-step cases and semi-infinite cases elsewhere.
The Casoratian identities imply equivalences between the deformed systems
obtained by multi-step Darboux transformations in terms of pseudo virtual
state vectors and those in terms of eigenvectors with shifted parameters.

Curbera and Dur\'an studied similar Casoratian identities for the Charlier,
Meixner and Hahn polynomials \cite{cd16}, which have the sinusoidal
coordinate $\eta(x)=x$. Their method is based on the Krall discrete measure.
Let us consider the case $\mathcal{N}=\max(\mathcal{D})$ and
$\min(\mathcal{D})\geq1$.
Then their map $I$ implies $I(\mathcal{D})=\bar{\mathcal{D}}$ and
$I(\bar{\mathcal{D}})=\mathcal{D}$, and we have
$\max(\bar{\mathcal{D}})=\mathcal{N}$ and $\min(\bar{\mathcal{D}})\geq1$.
Our identities \eqref{casoidred} correspond to their Theorem 1.1, 5.1 and
7.1 as follows:
Charlier: $F=\bar{\mathcal{D}}$,
Meixner: $F_1=\bar{\mathcal{D}}$ and $F_2=\emptyset$,
Hahn: $F_1=\bar{\mathcal{D}}$ and $F_2=F_3=\emptyset$
(Remark: $I(\emptyset)=\emptyset$ and $\max(\emptyset)=-1$).
The proportionality constants are also presented.

Among the reduced form polynomials, the big $q$-Jacobi family and
the discrete $q$-Hermite \Romannumeral{2} are not mentioned in this paper.
The orthogonality relations of the big $q$-Jacobi family are expressed in terms
of the Jackson integral and their rdQM need two component formalism
\cite{os34}.
The rdQM for the discrete $q$-Hermite \Romannumeral{2} is an infinite system,
$x\in\mathbb{Z}$.
We have not completed the study of the pseudo virtual state vectors for
these two systems.
It is plausible that similar Casoratian identities, which are polynomial
identities, do exist.
In fact R.\,Sasaki has checked tentative Casoratian identities for the big
$q$-Jacobi family (private communication).

In \S\,\ref{sec:discsym} twist operations (\romannumeral1)--(\romannumeral2)
and
($\widetilde{\text{\romannumeral1}}$)--($\widetilde{\text{\romannumeral2}}$)
are presented.
There are more discrete symmetries (\romannumeral3)--(\romannumeral4) for R
and $q$R, and
($\widetilde{\text{\romannumeral3}}$)--($\widetilde{\text{\romannumeral4}}$)
for $q$R:
\begin{align}
  (\text{\romannumeral3}):&
  \ \ \mathfrak{t}(x)\eqdef-x-1,
  \ \ \mathfrak{t}(\bm{\lambda})\eqdef
  (1+\lambda_1-\lambda_4,1+\lambda_2-\lambda_4,2-\lambda_3,2-\lambda_4),
  \ \ \mathfrak{t}(q)\eqdef q,
  \label{twist(iii)}\\
  (\text{\romannumeral4}):&
  \ \ \mathfrak{t}(x)\eqdef x+\lambda_3-1,
  \ \ \mathfrak{t}(\bm{\lambda})\eqdef
  (1+\lambda_1-\lambda_3,1+\lambda_2-\lambda_3,2-\lambda_3,
  2-2\lambda_3+\lambda_4),\n
  &\ \ \mathfrak{t}(q)\eqdef q,
  \label{twist(iv)}\\
  (\widetilde{\text{\romannumeral3}}):&
  \ \ (\text{\romannumeral3})\text{ with the replacement }
  \mathfrak{t}(q)\eqdef q^{-1},
  \label{twist(iii)t}\\
  (\widetilde{\text{\romannumeral4}}):&
  \ \ (\text{\romannumeral4})\text{ with the replacement }
  \mathfrak{t}(q)\eqdef q^{-1},
  \label{twist(iv)t}
\end{align}
all of which give the relation
\begin{equation}
  \tilde{\mathcal{E}}_{\text{v}}(\bm{\lambda})
  =\mathcal{E}_{\text{v}+N+1}(\bm{\lambda}).
\end{equation}
It is an interesting problem to clarify whether these twists give new pseudo
virtual state vectors and Casoratian identities or not.

\section*{Acknowledgments}

I thank R.\,Sasaki for discussion and reading of the manuscript.

\bigskip
\appendix
\section{Data of Orthogonal Polynomials}
\label{app:data}

In this appendix we present some necessary data of the orthogonal polynomials
\cite{kls,os12}.
The data of the Askey-Wilson polynomial are given in \ref{app:AW}.
The data of the polynomials appearing in finite rdQM are given in
\ref{app:finite} and those for semi-infinite rdQM are given in
\ref{app:semiinfinite}.

\subsection{Askey-Wilson}
\label{app:AW}

The fundamental data of the Askey-Wilson polynomial are \cite{kls,os30}
\begin{align}
  &q^{\bm{\lambda}}=(a_1,a_2,a_3,a_4)
  \ \ \bigl(\{a_1^*,a_2^*,a_3^*,a_4^*\}=\{a_1,a_2,a_3,a_4\}
  \text{ as a set}\bigr),\quad
  \gamma=\log q,\n
  &\bm{\delta}=(\tfrac12,\tfrac12,\tfrac12,\tfrac12),\quad\kappa=q^{-1},\quad
  \mathcal{E}_n(\bm{\lambda})=(q^{-n}-1)(1-b_4q^{n-1}),\quad
  b_4\eqdef a_1a_2a_3a_4,\n
  &\eta(x)=\cos x,\quad\varphi(x)=2\sin x,\n
  &\check{P}_n(x;\bm{\lambda})=P_n\bigl(\eta(x);\bm{\lambda}\bigr)
  =p_n\bigl(\eta(x);a_1,a_2,a_3,a_4|q\bigr)
  \label{AW}\\
  &\phantom{\check{P}_n(x;\bm{\lambda})}
  =a_1^{-n}(a_1a_2,a_1a_3,a_1a_4\,;q)_n\,
  {}_4\phi_3\Bigl(\genfrac{}{}{0pt}{}{q^{-n},\,b_4q^{n-1},\,
  a_1e^{ix},\,a_1e^{-ix}}{a_1a_2,\,a_1a_3,\,a_1a_4}\!\!\Bigm|\!q\,;q\Bigr),\n
  &V(x;\bm{\lambda})=\frac{\prod_{j=1}^4(1-a_je^{ix})}
  {(1-e^{2ix})(1-qe^{2ix})},\quad
  V^*(x;\bm{\lambda})=\frac{\prod_{j=1}^4(1-a_je^{-ix})}
  {(1-e^{-2ix})(1-qe^{-2ix})},\nonumber
\end{align}
where $p_n(\eta;a_1,a_2,a_3,a_4|q)$ is the Askey-Wilson polynomial.
Note that the Askey-Wilson system is invariant under the permutation
of $a_j$'s.

The pseudo virtual state wavefunction is defined by using the twist
operation, the discrete symmetry of the Hamiltonian \cite{os30}.
The twist operation $\mathfrak{t}$ etc.\ are
\begin{align}
  &\mathfrak{t}(\bm{\lambda})
  \eqdef(1-\lambda_1,1-\lambda_2,1-\lambda_3,1-\lambda_4)
  =2\bm{\delta}-\bm{\lambda},
  \label{twistAW}\\
  &V'(x;\bm{\lambda})\eqdef V\bigl(x;\mathfrak{t}(\bm{\lambda})\bigr),\quad
  \alpha(\bm{\lambda})=b_4q^{-2},\quad
  \alpha'(\bm{\lambda})=-(1-q)(1-b_4q^{-2}),
  \label{V'}\\
  &\mathcal{H}(\bm{\lambda})=\alpha(\bm{\lambda})\mathcal{H}'(\bm{\lambda})
  +\alpha'(\bm{\lambda}),\quad
\tilde{\mathcal{E}}_{\text{v}}(\bm{\lambda})\eqdef
  \alpha(\bm{\lambda})
  \mathcal{E}_{\text{v}}\bigl(\mathfrak{t}(\bm{\lambda})\bigr)
  +\alpha'(\bm{\lambda})
  =\mathcal{E}_{-\text{v}-1}(\bm{\lambda}).
  \label{EtvAW}
\end{align}
The pseudo virtual state polynomial $\check{\xi}_{\text{v}}(x;\bm{\lambda})
=\xi_{\text{v}}\bigl(\eta(x);\bm{\lambda}\bigr)$
is defined by
\begin{equation}
  \check{\xi}_{\text{v}}(x;\bm{\lambda})
  \eqdef\check{P}_{\text{v}}\bigl(x;\mathfrak{t}(\bm{\lambda})\bigr).
  \label{xivAW}
\end{equation}
The auxiliary function $\varphi_M(x)$ ($M\in\mathbb{Z}_{\geq 0}$)
is defined by
\begin{align}
  \varphi_M(x)&\eqdef
  \varphi(x)^{[\frac{M}{2}]}\prod_{k=1}^{M-2}
  \bigl(\varphi(x-i\tfrac{k}{2}\gamma)\varphi(x+i\tfrac{k}{2}\gamma)
  \bigr)^{[\frac{M-k}{2}]}\n
  &=\prod_{1\leq j<k\leq M}
  \frac{\eta\bigl(x+i(\frac{M+1}{2}-j)\gamma\bigr)
  -\eta\bigl(x+i(\frac{M+1}{2}-k)\gamma\bigr)}
  {\varphi(i\frac{j}{2}\gamma)}
  \times(-2)^{\frac12M(M-1)},
  \label{varphiMAW}
\end{align}
and $\varphi_0(x)=\varphi_1(x)=1$.
Here $[x]$ denotes the greatest integer not exceeding $x$.

\subsection{Finite cases}
\label{app:finite}

Data of the polynomials appearing in finite rdQM are presented \cite{kls,os12}.
Although there are two possible parameter choices indexed by
$\epsilon=\pm 1$ for the Hahn, dual Hahn, $q$-Hahn and dual $q$-Hahn
polynomials, we take $\epsilon=1$ for simplicity of presentation.

\subsubsection{Hahn (Ha)}
\label{app:Ha}

\begin{align}
  &\bm{\lambda}=(a,b,N),\quad
  \bm{\delta}=(1,1,-1),\quad\kappa=1,\quad
  \mathcal{E}_n(\bm{\lambda})=n(n+a+b-1),\n
  &\eta(x)=x,\quad
  \varphi(x)=1,\n
  &\check{P}_n(x;\bm{\lambda})=
  Q_n\bigl(\eta(x);a-1,b-1,N\bigr)
  ={}_3F_2\Bigl(
  \genfrac{}{}{0pt}{}{-n,\,n+a+b-1,\,-x}{a,\,-N}\Bigm|1\Bigr),\\
  &B(x;\bm{\lambda})=(x+a)(N-x),\quad
  D(x;\bm{\lambda})=x(b+N-x),\nonumber
\end{align}
where $Q_n(\eta;\alpha,\beta,N)$ is the Hahn polynomial in the conventional
parametrization \cite{kls}.
The Hahn polynomial is obtained from the Racah polynomial by
\begin{equation}
  \bm{\lambda}^{\text{R}}=(a,b+N+d,-N,d),\ \ d\to\infty:
  \ \ \lim_{d\to\infty}\check{P}^{\text{R}}_n(x;\bm{\lambda}^{\text{R}})
  =\check{P}_n(x;\bm{\lambda}).
\end{equation}

\subsubsection{dual Hahn (dHa)}
\label{app:dHa}

\begin{align}
  &\bm{\lambda}=(a,b,N),\quad
  \bm{\delta}=(1,0,-1),\quad\kappa=1,\quad
  \mathcal{E}_n=n,\n
  &\eta(x;\bm{\lambda})=x(x+a+b-1),\quad
  \varphi(x;\bm{\lambda})=\frac{2x+a+b}{a+b},\n
  &\check{P}_n(x;\bm{\lambda})=
  R_n\bigl(\eta(x;\bm{\lambda});a-1,b-1,N\bigr)
  ={}_3F_2\Bigl(
  \genfrac{}{}{0pt}{}{-n,\,x+a+b-1,\,-x}{a,\,-N}\Bigm|1\Bigr),\\
  &B(x;\bm{\lambda})=\frac{(x+a)(x+a+b-1)(N-x)}{(2x-1+a+b)(2x+a+b)},\quad
  D(x;\bm{\lambda})=\frac{x(x+b-1)(x+a+b+N-1)}{(2x-2+a+b)(2x-1+a+b)},\nonumber
\end{align}
where $R_n(\eta;\gamma,\delta,N)$ is the dual Hahn polynomial in the
conventional parametrization \cite{kls}.
The dual Hahn polynomial is obtained from the Racah polynomial by
\begin{equation}
  \bm{\lambda}^{\text{R}}=(a,b',-N,a+b-1),\ \ b'\to\infty:
  \ \ \lim_{b'\to\infty}\check{P}^{\text{R}}_n(x;\bm{\lambda}^{\text{R}})
  =\check{P}_n(x;\bm{\lambda}).
\end{equation}

\subsubsection{Krawtchouk (K)}
\label{app:K}

\begin{align}
  &\bm{\lambda}=(p,N),\quad
  \bm{\delta}=(0,-1),\quad \kappa=1,\quad
  \mathcal{E}_n=n,\quad
  \eta(x)=x,\quad\varphi(x)=1,\n
  &\check{P}_n(x;\bm{\lambda})
  =K_n\bigl(\eta(x);p,N\bigr)
  ={}_2F_1\Bigl(
  \genfrac{}{}{0pt}{}{-n,\,-x}{-N}\Bigm|p^{-1}\Bigr),\\
  &B(x;\bm{\lambda})=p(N-x),\quad
  D(x;\bm{\lambda})=(1-p)x,\nonumber
\end{align}
where $K_n(\eta;p,N)$ is the Krawtchouk polynomial.
The Krawtchouk polynomial is obtained from the Hahn polynomial by
\begin{equation}
  \bm{\lambda}^{\text{Ha}}=(1+pt,1+(1-p)t,N),\ \ t\to\infty:
  \ \ \lim_{t\to\infty}\check{P}^{\text{Ha}}_n(x;\bm{\lambda}^{\text{Ha}})
  =\check{P}_n(x;\bm{\lambda}).
\end{equation}

\subsubsection{$q$-Hahn ($q$Ha)}
\label{app:qHa}

\begin{align}
  &q^{\bm{\lambda}}=(a,b,q^N),\quad
  \bm{\delta}=(1,1,-1),\quad \kappa=q^{-1},\quad
  \mathcal{E}_n(\bm{\lambda})=(q^{-n}-1)(1-abq^{n-1}),\n
  &\eta(x)=q^{-x}-1,\quad
  \varphi(x)=q^{-x},\n
  &\check{P}_n(x;\bm{\lambda})
  =Q_n\bigl(1+\eta(x);aq^{-1},bq^{-1},N|q\bigr)
  ={}_3\phi_2\Bigl(
  \genfrac{}{}{0pt}{}{q^{-n},\,abq^{n-1},\,q^{-x}}
  {a,\,q^{-N}}\Bigm|q\,;q\Bigr),\\
  &B(x;\bm{\lambda})=(1-aq^x)(q^{x-N}-1),\quad
  D(x;\bm{\lambda})=aq^{-1}(1-q^x)(q^{x-N}-b),\nonumber
\end{align}
where $Q_n(\eta;\alpha,\beta,N|q)$ is the $q$-Hahn polynomial in the
conventional parametrization \cite{kls}.
The $q$-Hahn polynomial is obtained from the $q$-Racah polynomial by
\begin{equation}
  q^{\bm{\lambda}^{\text{$q$R}}}=(a,bq^Nd,q^{-N},d),\ \ d\to 0:
  \ \ \lim_{d\to 0}\check{P}^{\text{$q$R}}_n(x;\bm{\lambda}^{\text{$q$R}})
  =\check{P}_n(x;\bm{\lambda}).
\end{equation}

\subsubsection{dual $q$-Hahn (d$q$Ha)}
\label{app:dqHa}

\begin{align}
  &q^{\bm{\lambda}}=(a,b,q^N),\quad
  \bm{\delta}=(1,0,-1),\quad \kappa=q^{-1},\quad
  \mathcal{E}_n=q^{-n}-1,\n
  &\eta(x;\bm{\lambda})=(q^{-x}-1)(1-abq^{x-1}),\quad
  \varphi(x;\bm{\lambda})=\frac{q^{-x}-abq^x}{1-ab},\n
  &\check{P}_n(x;\bm{\lambda})
  =R_n\bigl(1+abq^{-1}+\eta(x;\bm{\lambda});aq^{-1},bq^{-1},N|q\bigr)
  ={}_3\phi_2\Bigl(
  \genfrac{}{}{0pt}{}{q^{-n},\,abq^{x-1},\,q^{-x}}
  {a,\,q^{-N}}\Bigm|q\,;q\Bigr),\!\!\\
  &B(x;\bm{\lambda})=
  \frac{(q^{x-N}-1)(1-aq^x)(1-abq^{x-1})}
  {(1-abq^{2x-1})(1-abq^{2x})},\n
  &D(x;\bm{\lambda})=aq^{x-N-1}
  \frac{(1-q^x)(1-abq^{x+N-1})(1-bq^{x-1})}
  {(1-abq^{2x-2})(1-abq^{2x-1})},\nonumber
\end{align}
where $R_n(\eta;\gamma,\delta,N|q)$ is the dual $q$-Hahn polynomial in the
conventional parametrization \cite{kls}.
The dual $q$-Hahn polynomial is obtained from the $q$-Racah polynomial by
\begin{equation}
  q^{\bm{\lambda}^{\text{$q$R}}}=(a,b',q^{-N},abq^{-1}),\ \ b'\to 0:
  \ \ \lim_{b'\to 0}\check{P}^{\text{$q$R}}_n(x;\bm{\lambda}^{\text{$q$R}})
  =\check{P}_n(x;\bm{\lambda}).
\end{equation}

\subsubsection{quantum $q$-Krawtchouk (q$q$K)}
\label{app:qqK}

\begin{align}
  &q^{\bm{\lambda}}=(p,q^N),\quad
  \bm{\delta}=(1,-1),\quad \kappa=q,\quad
  \mathcal{E}_n=1-q^n,\quad
  \eta(x)=q^{-x}-1,\quad
  \varphi(x)=q^{-x},\n
  &\check{P}_n(x;\bm{\lambda})
  =K^{\text{qtm}}_n\bigl(1+\eta(x);p,N;q\bigr)
  ={}_2\phi_1\Bigl(
  \genfrac{}{}{0pt}{}{q^{-n},\,q^{-x}}{q^{-N}}\Bigm|q\,;pq^{n+1}\Bigr),\\
  &B(x;\bm{\lambda})=p^{-1}q^x(q^{x-N}-1),\quad
  D(x;\bm{\lambda})=(1-q^x)(1-p^{-1}q^{x-N-1}),\nonumber
\end{align}
where $K^{\text{qtm}}_n(\eta;p,N;q)$ is the quantum $q$-Krawtchouk polynomial.
The quantum $q$-Krawtchouk polynomial is obtained from the $q$-Hahn polynomial
by
\begin{equation}
  q^{\bm{\lambda}^{\text{$q$Ha}}}=(a,pq,q^N),\ \ a\to\infty:
  \ \ \lim_{a\to\infty}\check{P}^{\text{$q$Ha}}_n
  (x;\bm{\lambda}^{\text{$q$Ha}})
  =\check{P}_n(x;\bm{\lambda}).
\end{equation}

\subsubsection{$q$-Krawtchouk ($q$K)}
\label{app:qK}

\begin{align}
  &q^{\bm{\lambda}}=(p,q^N),\
  \bm{\delta}=(2,-1),\quad \kappa=q^{-1},\quad
  \mathcal{E}_n(\bm{\lambda})=(q^{-n}-1)(1+pq^n),\n
  &\eta(x)=q^{-x}-1,\quad
  \varphi(x)=q^{-x},\n
  &\check{P}_n(x;\bm{\lambda})
  =K_n\bigl(1+\eta(x);p,N;q\bigr)
  ={}_3\phi_2\Bigl(
  \genfrac{}{}{0pt}{}{q^{-n},\,q^{-x},\,-pq^n}{q^{-N},\,0}\Bigm|q\,;q\Bigr),\\
  &B(x;\bm{\lambda})=q^{x-N}-1,\quad
  D(x;\bm{\lambda})=p(1-q^x),\nonumber
\end{align}
where $K_n(\eta;p,N;q)$ is the $q$-Krawtchouk polynomial.
The $q$-Krawtchouk polynomial is obtained from the $q$-Hahn polynomial by
\begin{equation}
  q^{\bm{\lambda}^{\text{$q$Ha}}}=(a,-a^{-1}pq,q^N),\ \ a\to 0:
  \ \ \lim_{a\to 0}\check{P}^{\text{$q$Ha}}_n(x;\bm{\lambda}^{\text{$q$Ha}})
  =\check{P}_n(x;\bm{\lambda}).
\end{equation}

\subsubsection{dual $q$-Krawtchouk (d$q$K)}
\label{app:dqK}

\begin{align}
  &q^{\bm{\lambda}}=(c,q^N),\quad
  \bm{\delta}=(0,-1),\quad \kappa=q^{-1},\quad
  \mathcal{E}_n=q^{-n}-1,\n
  &\eta(x;\bm{\lambda})=(q^{-x}-1)(1-cq^{x-N}),\quad
  \varphi(x;\bm{\lambda})=\frac{q^{-x}-cq^{1-N}q^x}{1-cq^{1-N}},\n
  &\check{P}_n(x;\bm{\lambda})
  =K_n\bigl(1+cq^{-N}+\eta(x;\bm{\lambda});c,N|q\bigr)
  ={}_3\phi_2\Bigl(
  \genfrac{}{}{0pt}{}{q^{-n},\,q^{-x},\,cq^{x-N}}{q^{-N},\,0}\Bigm|
  q\,;q\Bigr),\\
  &B(x;\bm{\lambda})=\frac{(q^{x-N}-1)(1-cq^{x-N})}
  {(1-cq^{2x-N})(1-cq^{2x+1-N})},
  \ \ D(x;\bm{\lambda})=-cq^{2x-2N-1}\frac{(1-q^x)(1-cq^x)}
  {(1-cq^{2x-1-N})(1-cq^{2x-N})},\nonumber
\end{align}
where $K_n(\eta;c,N|q)$ is the dual $q$-Krawtchouk polynomial in the
conventional parametrization \cite{kls}.
The dual $q$-Krawtchouk polynomial is obtained from the dual $q$-Hahn
polynomial by
\begin{equation}
  q^{\bm{\lambda}^{\text{d$q$Ha}}}=(a,a^{-1}cq^{1-N},q^N),\ \ a\to 0:
  \ \ \lim_{a\to 0}\check{P}^{\text{d$q$Ha}}_n(x;\bm{\lambda}^{\text{d$q$Ha}})
  =\check{P}_n(x;\bm{\lambda}).
\end{equation}

\subsubsection{affine $q$-Krawtchouk (a$q$K)}
\label{app:aqK}

\begin{align}
  &q^{\bm{\lambda}}=(p,q^N),\quad
  \bm{\delta}=(1,-1),\quad \kappa=q^{-1},\quad
  \mathcal{E}_n=q^{-n}-1,\quad
  \eta(x)=q^{-x}-1,\quad
  \varphi(x)=q^{-x},\n
  &\check{P}_n(x;\bm{\lambda})
  =K^{\text{aff}}_n\bigl(1+\eta(x);p,N;q\bigr)
  ={}_3\phi_2\Bigl(
  \genfrac{}{}{0pt}{}{q^{-n},\,q^{-x},\,0}{pq,\,q^{-N}}\Bigm|q\,;q\Bigr),\\
  &B(x;\bm{\lambda})=(q^{x-N}-1)(1-pq^{x+1}),\quad
  D(x;\bm{\lambda})=pq^{x-N}(1-q^x),\nonumber
\end{align}
where $K^{\text{aff}}_n(\eta;p,N;q)$ is the affine $q$-Krawtchouk polynomial.
The affine $q$-Krawtchouk polynomial is obtained from the $q$-Hahn polynomial
by
\begin{equation}
  q^{\bm{\lambda}^{\text{$q$Ha}}}=(pq,b,q^N),\ \ b\to 0:
  \ \ \lim_{b\to 0}\check{P}^{\text{$q$Ha}}_n(x;\bm{\lambda}^{\text{$q$Ha}})
  =\check{P}_n(x;\bm{\lambda}).
\end{equation}

\subsection{Semi-infinite cases}
\label{app:semiinfinite}

Data of the polynomials appearing in semi-infinite rdQM are presented
\cite{kls,os12}.

\subsubsection{Meixner (M)}
\label{app:M}

\begin{align}
  &\bm{\lambda}=(\beta,c),\quad
  \bm{\delta}=(1,0),\quad \kappa=1,\quad
  \mathcal{E}_n=n,\quad
  \eta(x)=x,\quad
  \varphi(x)=1,\n
  &\check{P}_n(x;\bm{\lambda})
  =M_n\bigl(\eta(x);\beta,c\bigr)
  ={}_2F_1\Bigl(
  \genfrac{}{}{0pt}{}{-n,\,-x}{\beta}\Bigm|1-c^{-1}\Bigr),\\
  &B(x;\bm{\lambda})=\frac{c}{1-c}(x+\beta),\quad
  D(x;\bm{\lambda})=\frac{1}{1-c}\,x,\nonumber
\end{align}
where $M_n(\eta;\beta,c)$ is the Meixner polynomial.
The Meixner polynomial is obtained from the Hahn polynomial by
\begin{equation}
  \bm{\lambda}^{\text{Ha}}=(\beta,1+\frac{1-c}{c}N,N),\ \ N\to\infty:
  \ \ \lim_{N\to\infty}\check{P}^{\text{Ha}}_n(x;\bm{\lambda}^{\text{Ha}})
  =\check{P}_n(x;\bm{\lambda}).
\end{equation}

\subsubsection{Charlier (C)}
\label{app:C}

\begin{align}
  &\bm{\lambda}=a,\quad
  \bm{\delta}=0,\quad \kappa=1,\quad
  \mathcal{E}_n=n,\quad
  \eta(x)=x,\quad
  \varphi(x)=1,\n
  &\check{P}_n(x;\bm{\lambda})
  =C_n\bigl(\eta(x);a)
  ={}_2F_0\Bigl(
  \genfrac{}{}{0pt}{}{-n,\,-x}{-}\Bigm|-a^{-1}\Bigr),\\
  &B(x;\bm{\lambda})=a,\quad
  D(x)=x,\nonumber
\end{align}
where $C_n(\eta;a)$ is the Charlier polynomial.
The Charlier polynomial is obtained from the Meixner polynomial by
\begin{equation}
  \bm{\lambda}^{\text{M}}=(\beta,\frac{a}{a+\beta}),\ \ \beta\to\infty:
  \ \ \lim_{\beta\to\infty}\check{P}^{\text{M}}_n(x;\bm{\lambda}^{\text{M}})
  =\check{P}_n(x;\bm{\lambda}).
\end{equation}

\subsubsection{little $q$-Jacobi (l$q$J)}
\label{app:lqJ}

\begin{align}
  &q^{\bm{\lambda}}=(a,b),\quad
  \bm{\delta}=(1,1),\quad \kappa=q^{-1},\quad
  \mathcal{E}_n(\bm{\lambda})=(q^{-n}-1)(1-abq^{n+1}),\n
  &\eta(x)=1-q^x,\quad
  \varphi(x)=q^x,\n
  &\check{P}_n(x;\bm{\lambda})
  =(-a)^{-n}q^{-\frac12n(n+1)}\frac{(aq\,;q)_n}{(bq\,;q)_n}\,
  p_n\bigl(1-\eta(x);a,b|q\bigr)\n
  &\phantom{\check{P}_n(x;\bm{\lambda})}
  =(-a)^{-n}q^{-\frac12n(n+1)}\frac{(aq;q)_n}{(bq;q)_n}\,
  {}_2\phi_1\Bigl(
  \genfrac{}{}{0pt}{}{q^{-n},\,abq^{n+1}}{aq}\Bigm|q\,;q^{x+1}\Bigr)\\
  &\phantom{\check{P}_n(x;\bm{\lambda})}
  ={}_3\phi_1\Bigl(
  \genfrac{}{}{0pt}{}{q^{-n},\,abq^{n+1},\,q^{-x}}{bq}
  \Bigm|q\,;a^{-1}q^x\Bigr),\n
  &B(x;\bm{\lambda})=a(q^{-x}-bq),\quad
  D(x)=q^{-x}-1,\nonumber
\end{align}
where $p_n(\eta;a,b|q)$ is the little $q$-Jacobi polynomial in the
conventional parametrization \cite{kls}.
The little $q$-Jacobi polynomial is obtained from the $q$-Hahn polynomial by
\begin{align}
  &x^{\text{$q$Ha}}=N-x,
  \ \ q^{\bm{\lambda}^{\text{$q$Ha}}}=(aq,bq,q^N),\ \ N\to\infty:\n
  &\lim_{N\to\infty}\check{P}^{\text{$q$Ha}}_n(x^{\text{$q$Ha}};
  \bm{\lambda}^{\text{$q$Ha}})
  =(-a)^nq^{\frac12n(n+1)}\frac{(bq;q)_n}{(aq;q)_n}
  \check{P}_n(x;\bm{\lambda}).
\end{align}

\subsubsection{$q$-Meixner ($q$M)}
\label{app:qM}

\begin{align}
  &q^{\bm{\lambda}}=(b,c),\quad
  \bm{\delta}=(1,-1),\quad \kappa=q,\quad
  \mathcal{E}_n=1-q^n,\quad
  \eta(x)=q^{-x}-1,\quad
  \varphi(x)=q^{-x},\n
  &\check{P}_n(x;\bm{\lambda})
  =M_n\bigl(1+\eta(x);b,c;q\bigr)
  ={}_2\phi_1\Bigl(
  \genfrac{}{}{0pt}{}{q^{-n},\,q^{-x}}{bq}\Bigm|q\,;-c^{-1}q^{n+1}\Bigr),\\
  &B(x;\bm{\lambda})=cq^x(1-bq^{x+1}),\quad
  D(x;\bm{\lambda})=(1-q^x)(1+bcq^x),\nonumber
\end{align}
where $M_n(\eta;b,c;q)$ is the $q$-Meixner polynomial.
The $q$-Meixner polynomial is obtained from the $q$-Hahn polynomial by
\begin{equation}
  q^{\bm{\lambda}^{\text{$q$Ha}}}=(bq,-b^{-1}c^{-1}q^{-N},q^N),\ \ N\to\infty:
  \ \ \lim_{N\to\infty}\check{P}^{\text{$q$Ha}}_n
  (x;\bm{\lambda}^{\text{$q$Ha}})
  =\check{P}_n(x;\bm{\lambda}).
\end{equation}

\subsubsection{little $q$-Laguerre/Wall (l$q$L)}
\label{app:lqL}

\begin{align}
  &q^{\bm{\lambda}}=a,\quad
  \bm{\delta}=1,\quad \kappa=q^{-1},\quad
  \mathcal{E}_n=q^{-n}-1,\quad
  \eta(x)=1-q^x,\quad
  \varphi(x)=q^x,\n
  &\check{P}_n(x;\bm{\lambda})
  =(a^{-1}q^{-n};q)_n\,p_n\bigl(1-\eta(x);a|q\bigr)
  ={}_2\phi_0\Bigl(
  \genfrac{}{}{0pt}{}{q^{-n},\,q^{-x}}{-}\Bigm|q\,;a^{-1}q^x\Bigr),\\
  &B(x;\bm{\lambda})=aq^{-x},\quad
  D(x)=q^{-x}-1,\nonumber
\end{align}
where $p_n(\eta;a|q)$ is the little $q$-Laguerre polynomial in the
conventional parametrization \cite{kls}.
The little $q$-Laguerre polynomial is obtained from the little $q$-Jacobi
polynomial by
\begin{equation}
  q^{\bm{\lambda}^{\text{l$q$J}}}=(a,b),\ \ b\to 0:
  \ \ \lim_{b\to 0}\check{P}^{\text{l$q$J}}_n(x;\bm{\lambda}^{\text{l$q$J}})
  =\check{P}_n(x;\bm{\lambda}).
\end{equation}

\subsubsection{Al-Salam-Carlitz \Romannumeral{2} (ASCII)}
\label{app:ASCII}

\begin{align}
  &q^{\bm{\lambda}}=a,\quad
  \bm{\delta}=0,\quad \kappa=q,\quad
  \mathcal{E}_n=1-q^n,\quad
  \eta(x)=q^{-x}-1,\quad
  \varphi(x)=q^{-x},\n
  &\check{P}_n(x;\bm{\lambda})
  =(-a)^{-n}q^{\frac12n(n-1)}\,V^{(a)}_n\bigl(1+\eta(x);q\bigr)
  ={}_2\phi_0\Bigl(
  \genfrac{}{}{0pt}{}{q^{-n},\,q^{-x}}{-}\Bigm|q\,;a^{-1}q^n\Bigr),\\
  &B(x;\bm{\lambda})=aq^{2x+1},\quad
  D(x;\bm{\lambda})=(1-q^x)(1-aq^x),\nonumber
\end{align}
where $V^{(a)}(\eta;q)$ is the Al-Salam-Carlitz \Romannumeral{2} polynomial
in the conventional parametrization \cite{kls}.
The Al-Salam-Carlitz \Romannumeral{2} polynomial is obtained from the
$q$-Meixner polynomial by
\begin{equation}
  q^{\bm{\lambda}^{\text{$q$M}}}=(-ac^{-1},c),\ \ c\to 0:
  \ \ \lim_{c\to 0}\check{P}^{\text{$q$M}}_n(x;\bm{\lambda}^{\text{$q$M}})
  =\check{P}_n(x;\bm{\lambda}).
\end{equation}

\subsubsection{$q$-Bessel ($q$B) (alternative $q$-Charlier)}
\label{app:aqC}

\begin{align}
  &q^{\bm{\lambda}}=a,\quad
  \bm{\delta}=2,\quad \kappa=q^{-1},\quad
  \mathcal{E}_n(\bm{\lambda})=(q^{-n}-1)(1+aq^n),\quad
  \eta(x)=1-q^x,\quad
  \varphi(x)=q^x,\n
  &\check{P}_n(x;\bm{\lambda})
  =(-a)^{-n}q^{-n^2}\,y_n\bigl(1-\eta(x);a;q)
  =q^{nx}\,{}_2\phi_1\Bigl(
  \genfrac{}{}{0pt}{}{q^{-n},\,q^{-x}}{0}\Bigm|q\,;-a^{-1}q^{1-n}\Bigr)\\
  &\phantom{\check{P}_n(x;\bm{\lambda})}
  =(-a)^{-n}q^{-n^2}
  {}_2\phi_1\Bigl(
  \genfrac{}{}{0pt}{}{q^{-n},\,-aq^n}{0}\Bigm|q\,;q^{x+1}\Bigr)
  ={}_3\phi_0\Bigl(
  \genfrac{}{}{0pt}{}{q^{-n},\,-aq^n,\,q^{-x}}{-}\Bigm|q\,;-a^{-1}q^x\Bigr),\n
  &B(x;\bm{\lambda})=a,\quad
  D(x)=q^{-x}-1,\nonumber
\end{align}
where $y_n(\eta;a;q)$ is the $q$-Bessel polynomial
(the alternative $q$-Charlier polynomial $K_n(\eta;a;q)$) in the conventional
parametrization \cite{kls}.
The $q$-Bessel polynomial is obtained from the little $q$-Jacobi polynomial by
\begin{equation}
  q^{\bm{\lambda}^{\text{l$q$J}}}=(a',-aa^{\prime\,-1}q^{-1}),\ \ a'\to 0:
  \ \ \lim_{a'\to 0}\check{P}^{\text{l$q$J}}_n(x;\bm{\lambda}^{\text{l$q$J}})
  =\check{P}_n(x;\bm{\lambda}).
\end{equation}

\subsubsection{$q$-Charlier ($q$C)}
\label{app:qC}

\begin{align}
  &q^{\bm{\lambda}}=a,\quad
  \bm{\delta}=-1,\quad \kappa=q,\quad
  \mathcal{E}_n=1-q^n,\quad
  \eta(x)=q^{-x}-1,\quad
  \varphi(x)=q^{-x},\n
  &\check{P}_n(x;\bm{\lambda})
  =C_n\bigl(1+\eta(x);a;q)
  ={}_2\phi_1\Bigl(
  \genfrac{}{}{0pt}{}{q^{-n},\,q^{-x}}{0}\Bigm|q\,;-a^{-1}q^{n+1}\Bigr),\\
  &B(x;\bm{\lambda})=aq^x,\quad
  D(x)=1-q^x,\nonumber
\end{align}
where $C_n(\eta;a;q)$ is the $q$-Charlier polynomial.
The $q$-Charlier polynomial is obtained from the $q$-Meixner polynomial by
\begin{equation}
  q^{\bm{\lambda}^{\text{$q$M}}}=(b,a),\ \ b\to 0:
  \ \ \lim_{b\to 0}\check{P}^{\text{$q$M}}_n(x;\bm{\lambda}^{\text{$q$M}})
  =\check{P}_n(x;\bm{\lambda}).
\end{equation}

\section{Pseudo Virtual State Vectors and Deformed Systems}
\label{app:pvsv}

In this appendix we explain pseudo virtual state vectors and deformed
systems obtained by one-step Darboux transformation in terms of pseudo
virtual state vectors.

We illustrate them by taking $q$R twist (\romannumeral1) case as an example.
We consider the following parameter range:
\begin{equation}
  c=q^{-N},\quad 0<ac<d<1,\quad qd<b<1,
\end{equation}
for which the Hamiltonian is well-defined, namely real symmetric, and
the constant $\alpha(\bm{\lambda})$ is positive.
Potential functions $B'(x;\bm{\lambda})$ and $D'(x;\bm{\lambda})$ are given
by \eqref{B'D'(i)(ii)}. We restrict the parameter range,
\begin{equation}
  ac<dq,\quad b<q,\quad d<q^2.
\end{equation}
Then $B'(x;\bm{\lambda})$ and $D'(x;\bm{\lambda})$ satisfy
\begin{align}
  &B'(x;\bm{\lambda})>0\ \ (x=0,1,\ldots,N),\ \ B'(-1;\bm{\lambda})=0,\n
  &D'(x;\bm{\lambda})>0\ \ (x=0,1,\ldots,N),\ \ D'(N+1;\bm{\lambda})=0.
  \label{B'>0}
\end{align}
By further restricting the parameter range, we attain the positivity of the
pseudo virtual state polynomial \eqref{xi(i)qR} in the extended domain,
\begin{equation}
  \check{\xi}_{\text{v}}(x;\bm{\lambda})>0\ \ (x=-1,0,\ldots,N+1).
\end{equation}
The range of $\text{v}$ may be restricted.
After \eqref{phi0}, let us define $\tilde{\phi}_0(x;\bm{\lambda})$ by
\begin{equation}
  \tilde{\phi}_0(x;\bm{\lambda})\eqdef\prod_{y=0}^{x-1}
  \sqrt{\frac{B'(x;\bm{\lambda})}{D'(x+1;\bm{\lambda})}}>0,
  \label{phit0}
\end{equation}
which is an `almost' zero mode of $\mathcal{H}'$ \eqref{H'},
\begin{equation}
  \mathcal{H}'(\bm{\lambda})\tilde{\phi}_0(x;\bm{\lambda})
  =D'(0;\bm{\lambda})\tilde{\phi}_0(0;\bm{\lambda})\delta_{x0}
  +B'(N;\bm{\lambda})\tilde{\phi}_0(N;\bm{\lambda})\delta_{xN}.
  \label{H'phi'0}
\end{equation}
Explicitly it is
\begin{equation}
  \tilde{\phi}_0(x;\bm{\lambda})
  =\frac{1-dq^{2x}}{1-d}\frac{1}{q^x\phi_0(x;\bm{\lambda})}.
\end{equation}

We define the pseudo virtual state vector
$\tilde{\phi}_{\text{v}}(x;\bm{\lambda})$ as follows:
\begin{equation}
  \tilde{\phi}_{\text{v}}(x;\bm{\lambda})\eqdef
  \tilde{\phi}_0(x;\bm{\lambda})\check{\xi}_{\text{v}}(x;\bm{\lambda}).
  \label{phitv}
\end{equation}
This pseudo virtual state vector satisfies
\begin{align}
  \mathcal{H}(\bm{\lambda})\tilde{\phi}_{\text{v}}(x;\bm{\lambda})
  &=\tilde{\mathcal{E}}_{\text{v}}(\bm{\lambda})
  \tilde{\phi}_{\text{v}}(x;\bm{\lambda})
  +\alpha(\bm{\lambda})D'(0;\bm{\lambda})\tilde{\phi}_0(0;\bm{\lambda})
  \check{\xi}_{\text{v}}(-1;\bm{\lambda})\delta_{x0}\n
  &\phantom{
  =\tilde{\mathcal{E}}_n(\bm{\lambda})\tilde{\phi}_{\text{v}}(x;\bm{\lambda})}
  \ +\alpha(\bm{\lambda})B'(N;\bm{\lambda})\tilde{\phi}_0(N;\bm{\lambda})
  \check{\xi}_{\text{v}}(N+1;\bm{\lambda})\delta_{xN},
\end{align}
where the pseudo virtual state energy
$\tilde{\mathcal{E}}_{\text{v}}(\bm{\lambda})$ is defined by (we present it
for twists (\romannumeral2) and
($\widetilde{\text{\romannumeral1}}$)--($\widetilde{\text{\romannumeral2}}$)
also)
\begin{align}
  (\text{\romannumeral1}),(\text{\romannumeral2}):&
  \ \ \tilde{\mathcal{E}}_{\text{v}}(\bm{\lambda})\eqdef
  \alpha(\bm{\lambda})
  \mathcal{E}_{\text{v}}\bigl(\mathfrak{t}(\bm{\lambda})\bigr)
  +\alpha'(\bm{\lambda}),
  \label{Etv(i)}\\
  (\widetilde{\text{\romannumeral1}}),(\widetilde{\text{\romannumeral2}}):&
  \ \ \tilde{\mathcal{E}}_{\text{v}}(\bm{\lambda};q)\eqdef
  \alpha(\bm{\lambda};q)
  \mathcal{E}_{\text{v}}\bigl(\mathfrak{t}(\bm{\lambda});q^{-1}\bigr)
  +\alpha'(\bm{\lambda};q).
  \label{Etv(i)t}
\end{align}
Namely the pseudo virtual state vector almost satisfies the Schr\"odinger
equation, except for both boundaries $x=0$ and $x=N$.
This is in good contrast to the virtual state vector in rdQM \cite{os26},
which fails to satisfy the Schr\"odinger equation
at only one of the boundaries.
We remark that the pseudo virtual state energies
$\tilde{\mathcal{E}}_{\text{v}}(\bm{\lambda})$ for
(\romannumeral1)--(\romannumeral2) and
($\widetilde{\text{\romannumeral1}}$)--($\widetilde{\text{\romannumeral2}}$)
give the same value,
\begin{equation}
  \tilde{\mathcal{E}}_{\text{v}}(\bm{\lambda})
  =\mathcal{E}_{-\text{v}-1}(\bm{\lambda}),
  \label{Etv=E-v-1}
\end{equation}
which is important for the equivalence between the state adding and deleting
Darboux transformations \cite{os29,os30}.

Let us introduce potential functions $\hat{B}_{d_1}(x;\bm{\lambda})$
and $\hat{D}_{d_1}(x;\bm{\lambda})$ determined by one of the pseudo virtual
state polynomials $\check{\xi}_{d_1}(x;\bm{\lambda})$ :
\begin{equation}
  \hat{B}_{d_1}(x;\bm{\lambda})\eqdef\alpha(\bm{\lambda})B'(x;\bm{\lambda})
  \frac{\check{\xi}_{d_1}(x+1;\bm{\lambda})}
  {\check{\xi}_{d_1}(x;\bm{\lambda})},\quad
  \hat{D}_{d_1}(x;\bm{\lambda})\eqdef\alpha(\bm{\lambda})D'(x;\bm{\lambda})
  \frac{\check{\xi}_{d_1}(x-1;\bm{\lambda})}
  {\check{\xi}_{d_1}(x;\bm{\lambda})},
\end{equation}
which satisfy
\begin{align}
  &\hat{B}_{d_1}(x;\bm{\lambda})>0\ \ (x=0,1,\ldots,N),
  \ \ \hat{B}_{d_1}(-1;\bm{\lambda})=0,\n
  &\hat{D}_{d_1}(x;\bm{\lambda})>0\ \ (x=0,1,\ldots,N),
  \ \ \hat{D}_{d_1}(N+1;\bm{\lambda})=0,\\
  &B(x;\bm{\lambda})D(x+1;\bm{\lambda})
  =\hat{B}_{d_1}(x;\bm{\lambda})\hat{D}_{d_1}(x+1;\bm{\lambda}),\n
  &B(x;\bm{\lambda})+D(x;\bm{\lambda})
  =\hat{B}_{d_1}(x;\bm{\lambda})+\hat{D}_{d_1}(x;\bm{\lambda})
  +\tilde{\mathcal{E}}_{d_1}(\bm{\lambda}).
\end{align}
The original Hamiltonian $\mathcal{H}(\bm{\lambda})$ is rewritten by using
them:
\begin{align}
  \mathcal{H}(\bm{\lambda})&=
  -\sqrt{\hat{B}_{d_1}(x;\bm{\lambda})\hat{D}_{d_1}(x+1;\bm{\lambda})}
  \,e^{\partial}
  -\sqrt{\hat{B}_{d_1}(x-1;\bm{\lambda})\hat{D}_{d_1}(x;\bm{\lambda})}
  \,e^{-\partial}\n
  &\quad+\hat{B}_{d_1}(x;\bm{\lambda})+\hat{D}_{d_1}(x;\bm{\lambda})
  +\tilde{\mathcal{E}}_{d_1}(\bm{\lambda})\\[2pt]
  &\!\!\!\!\!\!\!\left(
  \begin{array}{l}
  =\Bigl(\sqrt{\hat{B}_{d_1}(x;\bm{\lambda})}
  -\sqrt{\hat{D}_{d_1}(x;\bm{\lambda})}\,e^{-\partial}\Bigr)
  \Bigl(\sqrt{\hat{B}_{d_1}(x;\bm{\lambda})}
  -e^{\partial}\sqrt{\hat{D}_{d_1}(x;\bm{\lambda})}\,\Bigr)\\[8pt]
  \quad
  +\hat{D}_{d_1}(0;\bm{\lambda})\delta_{x0}
  +\tilde{\mathcal{E}}_{d_1}(\bm{\lambda})
  \end{array}
  \right).\nonumber
\end{align}
For this $(N+1)\times(N+1)$ matrix
$\mathcal{H}=(\mathcal{H}_{x,y})_{0\leq x,y\leq N}$, we define a deformed
Hamiltonian $\mathcal{H}_{d_1}(\bm{\lambda})$, which is an $(N+2)\times(N+2)$
matrix $\mathcal{H}_{d_1}=(\mathcal{H}_{d_1\,x,y})_{-1\leq x,y\leq N}$,
\begin{align}
  \mathcal{H}_{d_1}(\bm{\lambda})&\eqdef
  -\sqrt{\hat{B}_{d_1}(x+1;\bm{\lambda})\hat{D}_{d_1}(x+1;\bm{\lambda})}
  \,e^{\partial}
  -\sqrt{\hat{B}_{d_1}(x;\bm{\lambda})\hat{D}_{d_1}(x;\bm{\lambda})}
  \,e^{-\partial}\n
  &\quad+\hat{B}_{d_1}(x;\bm{\lambda})+\hat{D}_{d_1}(x+1;\bm{\lambda})
  +\tilde{\mathcal{E}}_{d_1}(\bm{\lambda})\\[2pt]
  \biggl(&
  =\Bigl(\sqrt{\hat{B}_{d_1}(x;\bm{\lambda})}
  -e^{\partial}\sqrt{\hat{D}_{d_1}(x;\bm{\lambda})}\,\Bigr)
  \Bigl(\sqrt{\hat{B}_{d_1}(x;\bm{\lambda})}
  -\sqrt{\hat{D}_{d_1}(x;\bm{\lambda})}\,e^{-\partial}\Bigr)
  +\tilde{\mathcal{E}}_{d_1}(\bm{\lambda})
  \biggr).\nonumber
\end{align}
Note that, in the RHS, only $\hat{B}_{d_1}(x;\bm{\lambda})$ and
$\hat{D}_{d_1}(x;\bm{\lambda})$ for $x=0,1,\ldots,N$ appear due to
$\hat{B}_{d_1}(-1;\bm{\lambda})=\hat{D}_{d_1}(N+1;\bm{\lambda})=0$,
and $\check{\xi}_{d_1}(x;\bm{\lambda})$ for $x=-1,0,\ldots,N+1$ appears.
The deformed Hamiltonian $\mathcal{H}_{d_1}(\bm{\lambda})$ has $N+2$
eigenvectors; $N+1$ eigenvectors $\phi_{d_1\,n}(x;\bm{\lambda})$
($n=0,1,\ldots,N$) are inherited from the eigenvectors $\phi_n(x;\bm{\lambda})$
for the original Hamiltonian $\mathcal{H}(\bm{\lambda})$, and a new
eigenvector $\breve{\Phi}_{d_1;d_1}(x;\bm{\lambda})$ with the pseudo virtual
energy $\tilde{\mathcal{E}}_{d_1}(\bm{\lambda})$ is added:
\begin{align}
  \mathcal{H}_{d_1}(\bm{\lambda})\phi_{d_1\,n}(x;\bm{\lambda})
  &=\mathcal{E}_n(\bm{\lambda})\phi_{d_1\,n}(x;\bm{\lambda})
  \ \ (n=0,1,\ldots,N),\\
  \mathcal{H}_{d_1}(\bm{\lambda})\breve{\Phi}_{d_1;d_1}(x;\bm{\lambda})
  &=\tilde{\mathcal{E}}_{d_1}(\bm{\lambda})
  \breve{\Phi}_{d_1;d_1}(x;\bm{\lambda}).
\end{align}
Here $\phi_{d_1\,n}(x;\bm{\lambda})$ and
$\breve{\Phi}_{d_1;d_1}(x;\bm{\lambda})$ are given by
\begin{align}
  \phi_{d_1\,n}(x;\bm{\lambda})&\eqdef
  \sqrt{\hat{B}_{d_1}(x;\bm{\lambda})}\,\phi_n(x;\bm{\lambda})
  -\sqrt{\hat{D}_{d_1}(x+1;\bm{\lambda})}\,\phi_n(x+1;\bm{\lambda}),
  \label{phid1n}\\
  \breve{\Phi}_{d_1;d_1}(x;\bm{\lambda})&\eqdef
  \frac{\phi_0(x+1;\bm{\lambda}-\bm{\delta})}
  {\sqrt{\check{\xi}_{d_1}(x;\bm{\lambda})
  \check{\xi}_{d_1}(x+1;\bm{\lambda})}},
\end{align}
which are defined for $x=-1,0,\ldots,N$.
We remark that \eqref{phid1n} is rewritten as
\begin{equation}
  \phi_{d_1\,n}(x;\bm{\lambda})
  =\frac{-\sqrt{\alpha(\bm{\lambda})B'(x;\bm{\lambda})}\,
  \tilde{\phi}_0(x;\bm{\lambda})}
  {\sqrt{\check{\xi}_{d_1}(x;\bm{\lambda})
  \check{\xi}_{d_1}(x+1;\bm{\lambda})}}
  \text{W}_{\text{C}}[\check{\xi}_{d_1},\nu\check{P}_n](x;\bm{\lambda}),\quad
  \nu(x;\bm{\lambda})\eqdef
  \frac{\phi_0(x;\bm{\lambda})}{\tilde{\phi}_0(x;\bm{\lambda})}.
\end{equation}
This deformed Hamiltonian $\mathcal{H}_{d_1}(\bm{\lambda})$ can be
rewritten in the standard form:
\begin{align}
  \mathcal{H}_{d_1}(\bm{\lambda})&=
  -\sqrt{B_{d_1}(x;\bm{\lambda})D_{d_1}(x+1;\bm{\lambda})}
  \,e^{\partial}
  -\sqrt{B_{d_1}(x-1;\bm{\lambda})D_{d_1}(x;\bm{\lambda})}
  \,e^{-\partial}\n
  &\quad+B_{d_1}(x;\bm{\lambda})+D_{d_1}(x;\bm{\lambda})
  +\tilde{\mathcal{E}}_{d_1}(\bm{\lambda})\\
  &=\Bigl(\sqrt{B_{d_1}(x;\bm{\lambda})}
  -\sqrt{D_{d_1}(x;\bm{\lambda})}\,e^{-\partial}\Bigr)
  \Bigl(\sqrt{B_{d_1}(x;\bm{\lambda})}
  -e^{\partial}\sqrt{D_{d_1}(x;\bm{\lambda})}\,\Bigr)
  +\tilde{\mathcal{E}}_{d_1}(\bm{\lambda}).\nonumber
\end{align}
Here potential functions $B_{d_1}(x;\bm{\lambda})$ and
$D_{d_1}(x;\bm{\lambda})$ are
\begin{equation}
  B_{d_1}(x;\bm{\lambda})\eqdef\alpha(\bm{\lambda})D'(x+1;\bm{\lambda})
  \frac{\check{\xi}_{d_1}(x;\bm{\lambda})}
  {\check{\xi}_{d_1}(x+1;\bm{\lambda})},
  \ \ D_{d_1}(x;\bm{\lambda})\eqdef\alpha(\bm{\lambda})B'(x;\bm{\lambda})
  \frac{\check{\xi}_{d_1}(x+1;\bm{\lambda})}
  {\check{\xi}_{d_1}(x;\bm{\lambda})},
\end{equation}
which satisfy
\begin{align}
  &B_{d_1}(x;\bm{\lambda})>0\ \ (x=-1,0,\ldots,N-1),
  \ \ B_{d_1}(N;\bm{\lambda})=0,\n
  &D_{d_1}(x;\bm{\lambda})>0\ \ (x=0,1,\ldots,N),
  \ \ D_{d_1}(-1;\bm{\lambda})=0.
\end{align}

Repeating this deformation procedure $M$-times (the parameter range should
be restricted appropriately), we can obtain deformed Hamiltonians
$\mathcal{H}_{d_1\ldots d_M}(\bm{\lambda})$.
They are $(N+M+1)\times(N+M+1)$ matrices $\mathcal{H}_{d_1\ldots d_M}
=(\mathcal{H}_{d_1\ldots d_M\,x,y})_{-M\leq x,y\leq N}$ and their eigenvalues
are $\mathcal{E}_n(\bm{\lambda})$ ($n=0,1,\ldots,N$) and
$\tilde{\mathcal{E}}_{d_j}$ ($j=1,2,\ldots,M$).
We will report on this subject in detail elsewhere.

Next let us consider the equivalence between the above one-step deformation
in terms of the pseudo virtual state vector with $d_1=\ell$ and the
$\ell$-step deformation in terms of the eigenvectors with shifted parameters.
For simplicity we take $\mathcal{N}=\max(\mathcal{D})$.
The multi-step Darboux transformations in terms of the eigenvectors were
studied in \cite{os22}.
Simple examples, in which the eigenvectors $\phi_1,\phi_2,\ldots,\phi_{\ell}$
are deleted, are given in its Appendix A.
Eqs.\,(A.16), (A.24) and (A.26) in \cite{os22} for the $q$-Racah case:
\begin{equation*}
  \varphi_{\ell}(x;\bm{\lambda})^{-1}
  \text{W}_{\text{C}}[\check{P}_1,\check{P}_2,\ldots,\check{P}_{\ell}]
  (x;\bm{\lambda})
  \propto\check{P}_{\ell}\bigl(-x;-\bm{\lambda}-(\ell-1)\bm{\delta}),
\end{equation*}
namely,
\begin{equation}
  \check{\xi}_{\ell}(x-1;\bm{\lambda})\propto
  \varphi_{\ell}\bigl(x;\bm{\lambda}-(\ell+1)\bm{\delta}\bigr)^{-1}
  \text{W}_{\text{C}}[\check{P}_1,\check{P}_2,\ldots,\check{P}_{\ell}]
  \bigl(x;\bm{\lambda}-(\ell+1)\bm{\delta}\bigr).
\end{equation}
This is the special case of the Casoratian identities \eqref{casoidqR} with
$M=1$, $\mathcal{N}=\ell$, $\mathcal{D}=\{\ell\}$ and
$\bar{\mathcal{D}}=\{1,2,\ldots,\ell\}$.
Let us denote the potential functions of the deformed system in Appendix
A of \cite{os22} as
$B^{\text{KA}}_{\{1,2,\ldots,\ell\}}(x;\bm{\lambda})
=B^{\text{ref.\cite{os22}}}_{\ell}(x;\bm{\lambda})$ and
$D^{\text{KA}}_{\{1,2,\ldots,\ell\}}(x;\bm{\lambda})
=D^{\text{ref.\cite{os22}}}_{\ell}(x;\bm{\lambda})$.
The Hamiltonian of the deformed system in Appendix A of \cite{os22} is
\begin{align}
  \mathcal{H}^{\text{KA}}_{\{1,2,\ldots,\ell\}}(\bm{\lambda})
  &=\big(\mathcal{H}^{\text{KA}}_{\{1,2,\ldots,\ell\}\,;\,x,y}(\bm{\lambda})
  \bigr)_{0\leq x,y\leq N-\ell}\,,\n[2pt]
  \mathcal{H}^{\text{KA}}_{\{1,2,\ldots,\ell\}}(\bm{\lambda})&=
  -\sqrt{B^{\text{KA}}_{\{1,2,\ldots,\ell\}}(x;\bm{\lambda})
  D^{\text{KA}}_{\{1,2,\ldots,\ell\}}(x+1;\bm{\lambda})}
  \,e^{\partial}\\
  &\quad-\sqrt{B^{\text{KA}}_{\{1,2,\ldots,\ell\}}(x-1;\bm{\lambda})
  D^{\text{KA}}_{\{1,2,\ldots,\ell\}}(x;\bm{\lambda})}
  \,e^{-\partial}\n
  &\quad+B^{\text{KA}}_{\{1,2,\ldots,\ell\}}(x;\bm{\lambda})
  +D^{\text{KA}}_{\{1,2,\ldots,\ell\}}(x;\bm{\lambda}).
  \nonumber
\end{align}
Then eqs.\,(A.26), (A.29) and (A.30) in \cite{os22} give
\begin{align}
  \check{\xi}_{\ell}(x;\bm{\lambda})
  &=\check{\xi}^{\text{ref.\cite{os22}}}_{\ell}
  \bigl(x+1;\bm{\lambda}-(\ell+1)\bm{\delta}\bigr),\n
  B_{\ell}(x;\bm{\lambda})
  &=\kappa^{-\ell-1}B^{\text{KA}}_{\{1,2,\ldots,\ell\}}
  \bigl(x+1;\bm{\lambda}-(\ell+1)\bm{\delta}\bigr),\\
  D_{\ell}(x;\bm{\lambda})
  &=\kappa^{-\ell-1}D^{\text{KA}}_{\{1,2,\ldots,\ell\}}
  \bigl(x+1;\bm{\lambda}-(\ell+1)\bm{\delta}\bigr).\nonumber
\end{align}
Therefore we have
\begin{equation}
  \mathcal{H}_{\ell\,;\,x,y}(\bm{\lambda})
  -\tilde{\mathcal{E}}_{\ell}(\bm{\lambda})\delta_{x,y}
  =\kappa^{-\ell-1}\mathcal{H}^{\text{KA}}_{\{1,2,\ldots,\ell\}\,;\,x+1,y+1}
  \bigl(\bm{\lambda}-(\ell+1)\bm{\delta}\bigr),
\end{equation}
namely
\begin{equation}
  \Bigl(\mathcal{H}_{\ell\,;\,x,y}(\bm{\lambda})
  -\tilde{\mathcal{E}}_{\ell}(\bm{\lambda})\delta_{x,y}
  \Bigr)_{-1\leq x,y\leq N}
  =\kappa^{-\ell-1}\Bigl(
  \mathcal{H}^{\text{KA}}_{\{1,2,\ldots,\ell\}\,;\,x+1,y+1}
  \bigl(\bm{\lambda}-(\ell+1)\bm{\delta}\bigr)\Bigr)_{-1\leq x,y\leq N}.
\end{equation}
This establishes the equivalence of the two systems.


\end{document}